\documentclass[aps,prl,onecolumn,superscriptaddress,12pt]{revtex4}

\usepackage{graphicx}
\usepackage{color}
\usepackage{amsfonts}
\newcommand{\bise}{\,Bi$_2$Se$_3$}
\newcommand{\iA}{\, \AA$^{-1}$~}

\begin{document}
\title{The electronic structure of clean and adsorbate-covered \bise: an angle-resolved photoemission study}

\author{Marco~Bianchi}
\affiliation{Department of Physics and Astronomy, Interdisciplinary Nanoscience Center, Aarhus University, 8000 Aarhus C, Denmark}
\author{Richard~C.~Hatch}
\author{Dandan~Guan}
\affiliation{Department of Physics and Astronomy, Interdisciplinary Nanoscience Center, Aarhus University, 8000 Aarhus C, Denmark}
\author{Tilo Planke}
\affiliation{Department of Physics and Astronomy, Interdisciplinary Nanoscience Center, Aarhus University, 8000 Aarhus C, Denmark}
\author{Jianli Mi}
\affiliation{Center for Materials Crystallography, Department of Chemistry, Interdisciplinary Nanoscience Center, Aarhus University, 8000 Aarhus C, Denmark}
\author{Bo Brummerstedt Iversen}
\affiliation{Center for Materials Crystallography, Department of Chemistry, Interdisciplinary Nanoscience Center, Aarhus University, 8000 Aarhus C, Denmark}
\author{Philip~Hofmann}
\affiliation{Department of Physics and Astronomy, Interdisciplinary Nanoscience Center, Aarhus University, 8000 Aarhus C, Denmark}
\email[]{philip@phys.au.dk}

\date{\today}

\begin{abstract}
Angle-resolved photoelectron spectroscopy is used for a detailed study of the electronic structure of the topological insulator \bise. Nominally stoichiometric and calcium-doped samples were investigated. The pristine surface shows the topological surface state in the bulk band gap. As time passes, the Dirac point moves to higher binding energies, indicating an increasingly strong downward bending of the bands near the surface. This time-dependent band bending is related to a contamination of the surface and can be accelerated by intentionally exposing the surface to carbon monoxide and other species. For a sufficiently strong band bending, additional states appear at the Fermi level. These are interpreted as quantised conduction band states. For large band bendings, these states are found to undergo a strong Rashba splitting. The formation of quantum well states is also observed for the valence band states. Different interpretations of similar data are also discussed.
\end{abstract}


\maketitle

\section{Introduction}

The quantum Hall effect represents a state of matter in which an insulating, two-dimensional  sample permits perfect conductance along its edges \cite{Klitzing:1980}. This is an intriguing physical situation with many potential applications, but it can only be realized at low temperatures and in a high magnetic field. Since its discovery in 1980, researchers have searched for this perfect conductance in systems which do not have the constraints of magnetic field and temperature. Twenty five years later, it was suggested that a similar situation could be realized using materials with strong spin-orbit splitting of the states without the presence of a magnetic field \cite{Kane:2005b,Kane:2005c,Bernevig:2006b,Bernevig:2006,Moore:2007}. The corresponding state is called the quantum spin Hall effect (QSHE) and it was experimentally confirmed in 2007 \cite{Konig:2007}. In contrast to the quantum Hall effect, the QSHE supports counter-propagating states on each sample edge but the propagation direction is linked to the electron spin. Back-scattering is thus not permitted and, again, perfect one-dimensional transport is observed \cite{,Bernevig:2006b,Bernevig:2006,Konig:2007}.

While the QSHE corresponds to a two-dimensional system with one-dimensional edge states, its three-dimensional analogues are called topological insulators \cite{Zhang:2008,Moore:2010,Hasan:2010}. These materials are bulk band insulators but they support metallic surface states. The existence of such states is required by the bulk band structure, in contrast to the more coincidental and fragile metallic surface states frequently encountered on semiconductor surfaces  \cite{Barke:2007a}. The first bulk topological insulator found experimentally was obtained by doping the semimetal Bi with Sb to obtain Bi$_{1-x}$Sb$_{x}$ ($x \approx 0.1$) \cite{Fu:2007b,Murakami:2007,Hsieh:2008}, a material with a very small band gap, but stable surface states that, not surprisingly, strongly resemble those on the pure Bi surface \cite{Ast:2001}. As a material, Bi$_{1-x}$Sb$_{x}$ has several disadvantages for potential applications: its random nature (implying poor crystal quality), the very small bulk band gap of less than 30~meV which restricts possible applications to low temperatures, and the rather complicated electronic structure of its (111) surface. Later, other topological insulators were discovered such as Bi$_2$Se$_3$ and Bi$_2$Te$_3$ \cite{Zhang:2009,Xia:2009,Noh:2008,Chen:2009,Hsieh:2009c}. These materials are ordered alloys, have a gap size of $\approx300$~meV, and their surface electronic structure consists of a single, metallic state with a Dirac cone-like dispersion. 

The surface states on a three-dimensional topological insulator show strong similarity to the edge states of the QSHE. They are completely non-degenerate with respect to spin, except at some symmetry-protected points, and the spin texture of the Dirac cone shows a well-defined chirality of the in-plane spin polarisation  \cite{Hsieh:2009,Hsieh:2009b,Hsieh:2009c}. 
Consequently, the states are also immune to back-scattering \cite{Roushan:2009,Alpichshev:2010}.  Strictly spoken, however, this restriction applies to backscattering only, with near back-scattering processes  being unlikely but allowed. The situation on these surfaces is thus very similar to that for conventional surface states with strong Rashba-type spin-orbit splitting \cite{Petersen:2000,Nechaev:2009} and virtually the same as for the surface states on the pure semimetal Bi \cite{Gayone:2003,Hofmann:2006,Kim:2005b,Koroteev:2004,Pascual:2004}. 
 
\begin{figure}[htbp]
		\includegraphics[width=6in]{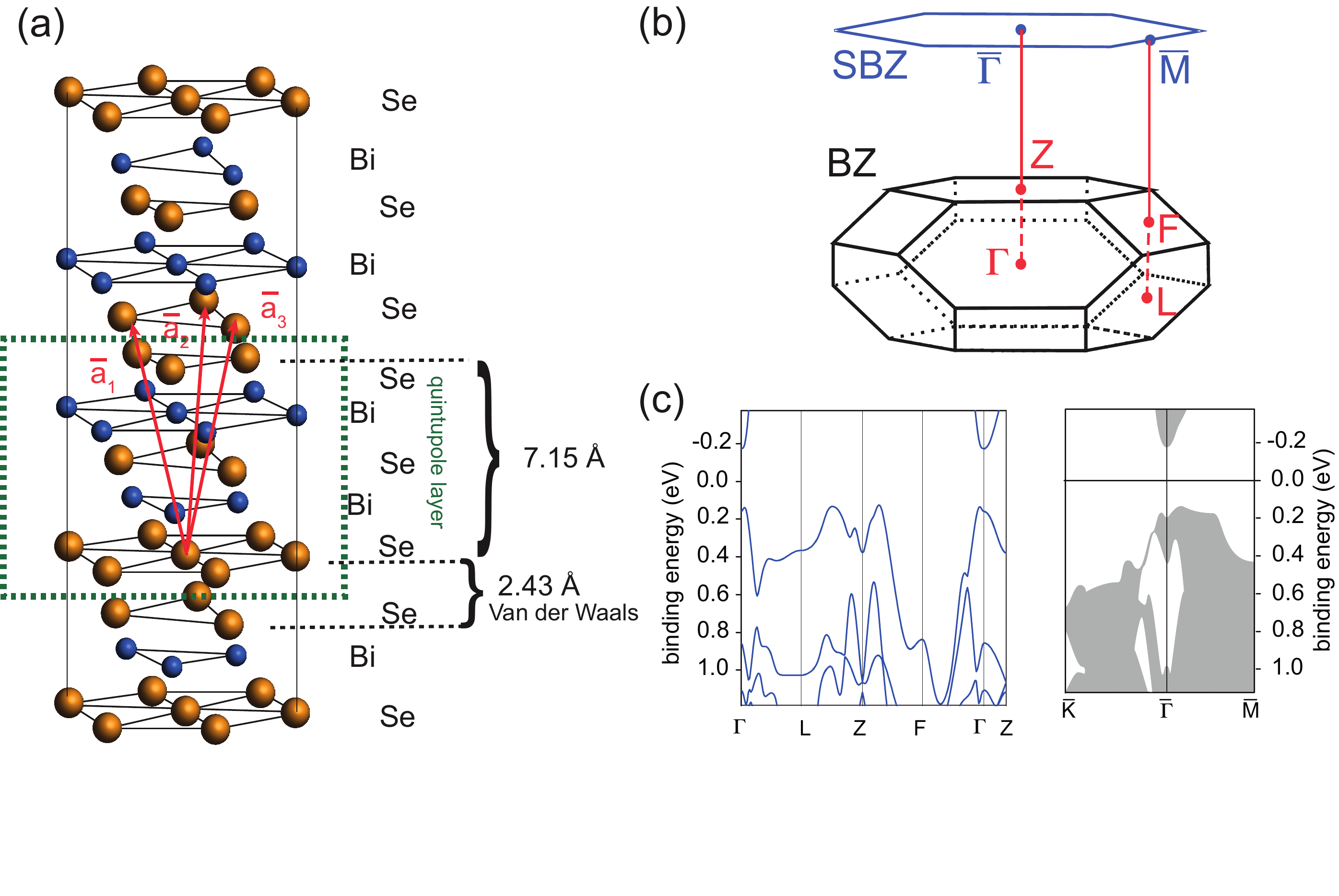}
		\caption{\footnotesize (a) Crystal structure of \bise~and interlayer distances as determined by X-ray diffraction. Three primitive lattice vectors are indicated in red while in green the quintuple layer is highlighted. (b) Bulk and surface Brillouin zones (BZ and SBZ, respectively) with bulk time-reversal invariant momenta (TRIMs) and their projection to surface TRIMs. (c) Bulk band structure along selected high symmetry points and projection on the (111) surface after Ref. \cite{Eremeev:2010b}. }
		\label{fig:BzCrystal}
\end{figure}

Recently, the binary alloys \bise~and Bi$_2$Te$_3$ \cite{Chen:2009} have been used as prototype topological insulators in experiments because of their simple electronic structure with a single Dirac cone at the surface, and because of practical matters such as ease of crystal fabrication and surface preparation. The latter is related to the layered structure of \bise, as shown in Fig. \ref{fig:BzCrystal}(a), that is formed of quintuple layers of covalently bonded \bise,~separated by van der Waals gaps. This crystal structure allows samples to be easily cleaved in vacuum thus exposing a clean surface.

The topological character of  \bise~is derived from a parity inversion at the $\Gamma$ point of  the bulk band structure  \cite{Zhang:2009}. To see this, consider the so-called time-reversal invariant momenta (TRIMs) $\mathbf{\Gamma_i}$  in the bulk Brillouin zone (BZ), as defined by  $-\mathbf{\Gamma}_i=\mathbf{\Gamma}_i+\mathbf{G}$, where $\mathbf{G}$ is a reciprocal lattice vector \cite{Teo:2008} (see Fig. \ref{fig:BzCrystal}(b)). For each TRIM, it is possible to define a parity invariant $\delta_i (\mathbf{\Gamma_i})$ that accounts for the parity of the occupied bands by taking the product of the parity eigenvalues of all (spin degenerate) bands at the TRIM. In \bise, these parity invariants are all $1$, except for the one corresponding to the bulk $\Gamma$ point which is $-1$   \cite{Zhang:2009}. Taking the product of the parity invariants at all 8 bulk TRIMs, one thus obtains again $-1$ and this guarantees the bulk material to be a strong topological insulator. These parity invariants are then projected onto the surface TRIMs (defined in an analogous way), and result in the so-called surface fermion parity. For the hexagonal surface Brillouin zone (SBZ) of \bise(111), the $\bar{\Gamma}$ surface TRIM thus obtains a surface fermion parity of $-1$ while the $\bar{M}$ TRIM has $1$. The surface electronic structure is then closely related to these projected parity eigenvalues, with a change in surface fermion parity requiring an odd number of Fermi level crossings between the TRIMS and, even more stringently, a surface fermion parity eigenvalue of $-1$  requiring the surface TRIM to be enclosed by an odd number of closed Fermi contours \cite{Teo:2008}. Consistent with this, the $\bar{\Gamma}$ point of \bise~ is enclosed by a single, circular Fermi contour.

Already in the first experimental studies of the topological surface states of \bise~by angle-resolved photoemission (ARPES), a peculiar ageing effect was reported in that the observed electronic states gradually move to higher binding energies \cite{Xia:2009,Hsieh:2009c}. 
This effect was earlier observed in the similar compound Bi$_2$Te$_3$ \cite{Noh:2008} and ascribed to band bending induced by a slow geometrical relaxation of the van der Waals gap between the quintuple layers of the structure. It turns out that this ageing effect brings about a number of interesting and controversial spectroscopic observations that will be discussed in this paper. 

The ageing-induced band bending shifts the observed states to higher binding energies such that the minimum of the bulk conduction band eventually becomes observable, even for initially p-doped samples \cite{Noh:2008,Hsieh:2009c}. Moreover, the spectroscopic appearance of the states near the conduction band changes. First, a distinct rim around the bottom of the conduction band becomes observable \cite{Bianchi:2010b}, then  this rim develops into a well-separated state, and eventually it shows a splitting that is strongly reminiscent of a Rashba-type spin-orbit splitting  \cite{Bychkov:1984b}  of two dimensional states, and is spectroscopically quite similar to Rashba-split surface states on surfaces involving heavy elements such as Au or Bi  \cite{Lashell:1996,Agergaard:2001,Koroteev:2004,Ast:2007}. Similar spectroscopic features were observed for surfaces which were exposed to magnetic atoms \cite{Wray:2011}, rest gas in the vacuum \cite{King:2011}, carbon monoxide \cite{Bianchi:2011}, water \cite{Benia:2011} and alkali atoms \cite{King:2011,Pan:2011a,Valla:2012}.

The observed rim around the conduction band was explained in terms of the formation of a quantum well state (or two-dimensional electron gas) in the quantum well which is formed because of the strong downward bending of the conduction band at the surface \cite{Bianchi:2010}. The Rashba-type splitting is a consequence of this quantum well state being placed in a strong electric field near the surface \cite{King:2011}. However, alternative explanations have also been given. The Rashba-split states have been discussed in terms of being additional, but topologically trivial, surface states whose presence becomes necessary because of the strong downward band bending \cite{Wray:2011}. Based on first principles calculations, it has also been suggested that the additional states could be caused by the expansion of near-surface van der Waals gaps due to the intercalation of adsorbates \cite{Eremeev:2011b,Menshchikova:2011,Ye:2011}.
While these calculations show that an increase in near-surface van der Waals gaps causes the electronic structure to become more two-dimensional, these predictions fail to explain all of the observed changes to the electronic structure as will be discussed in more detail below.

In the present paper we report ARPES results from clean \bise~surfaces, using crystals with different bulk doping. We also discuss carbon monoxide adsorption on Ca-doped samples. Special emphasis is given to the band-bending induced states in the region of the conduction band. The paper is structured as follows: we start with a brief review of the ARPES technique and discuss how states of different character (bulk, surface, two-dimensional quantum well) are identified in ARPES spectra. We then describe the experimental and theoretical tools used in this paper. This is followed by a section presenting and discussing the results as well as a brief conclusion.


\section{Angle-resolved photoemission}

Even though the ARPES technique is well known and established (for reviews and books see Refs. \cite{Himpsel:1985,Plummer:1982,Kevan:1992,Matzdorf:1998,Hufner:2003,Hofmann:2009b}), it is useful to note a few details that will be exploited in the experimental investigation presented in the following part. 
	
\begin{figure}[htbp]
		\includegraphics[width=4.5in]{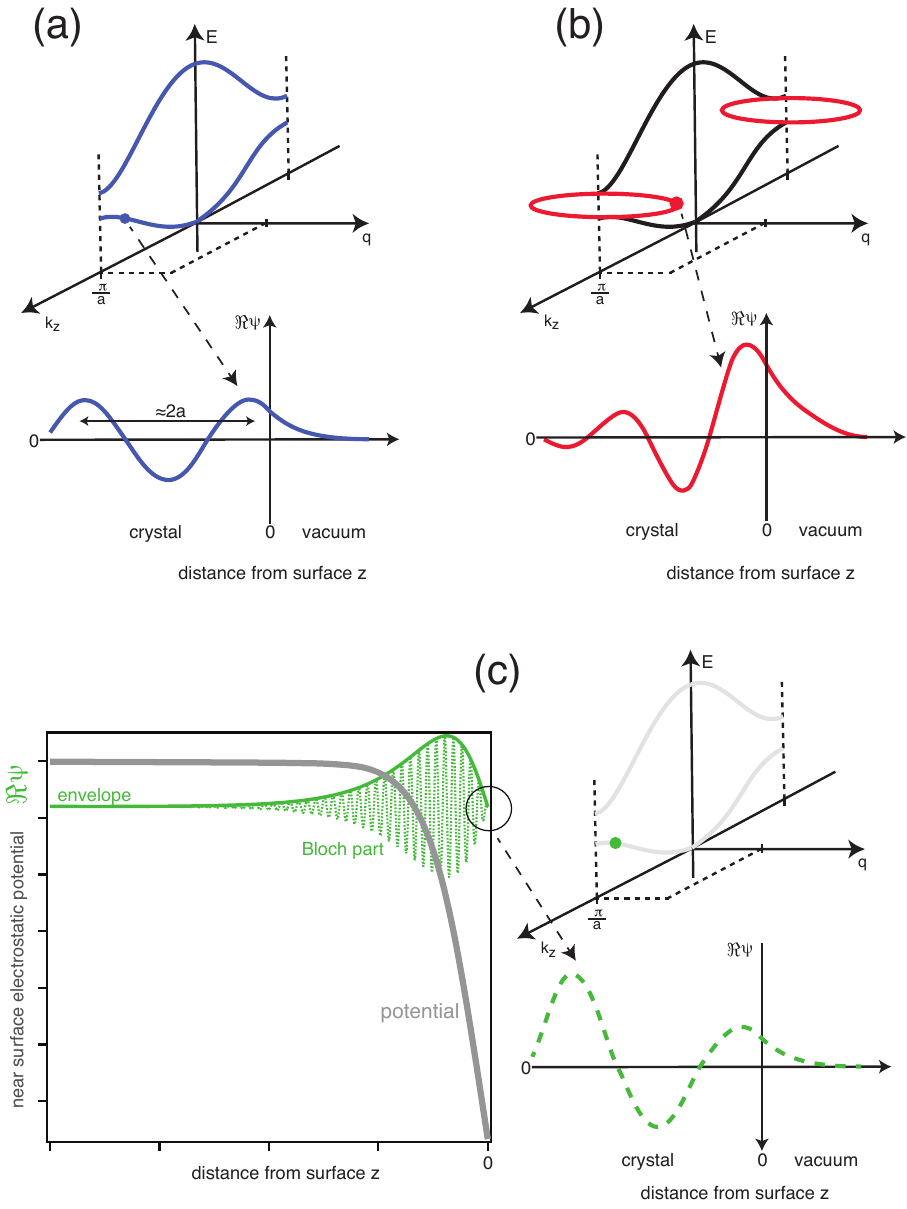}
		\caption{\footnotesize Electronic states near the surface of a solid. (a) Dispersion of nearly free-electron like bulk states and the wave function of such a state near the surface (real part). The one-dimensional lattice constant perpendicular to the surface is $a$. The Bloch wave inside the crystal can always be matched to an exponentially decaying tail outside the crystal. (b) (Projected) bulk band gaps can give rise to new Bloch states with a complex wave number $k_z + i q$, i.e. an exponential growth in one direction. In some cases, these exponentially increasing solutions can also be matched to an exponential tail outside the surface, giving rise to a surface state. (c) Left: In a semiconductor, a strong downward bending of the conduction band can lead to the quantisation of the conduction band states near the surface. The envelope wave function of the lowest state (solid line) has no node and the microscopic state still has Bloch-wave character (dashed line). Right: In a band structure picture, this quantisation reduces the permitted $k_z$-values in a band (only considered for the lower band). The states thus loose their three dimensional dispersion and become two-dimensional. The near-surface Bloch wave looks similar to that of the parent bulk state.}
		\label{fig:near_surface_states}
\end{figure}

We start by briefly reviewing the character of the electronic states that we can expect to detect at or near a surface. A schematic summary is given in Fig. \ref{fig:near_surface_states}. Bulk Bloch states (Fig. \ref{fig:near_surface_states}(a)) can always be matched to an exponentially decaying tail outside the surface by adjusting the phase between the incoming and reflected wave inside the solid and the amplitude of the evanescent wave outside. The bulk electronic structure is thus not significantly modified by the introduction of the surface and bulk states. Within (projected) band gaps, such as near the BZ boundary in the case of a nearly free electron model, new solutions to the bulk Schr\"odinger equation can exist. These give real energy eigenvalues, but only if a complex $k_z + i q$ wave vector is chosen. Clearly, such solutions cannot be normalized inside the bulk but at the surface it is in some cases also possible to match them to an exponentially decaying wave function in the vacuum. This gives rise to genuine surface states, a phenomenon that has been known for a long time \cite{Tamm:1932,Shockley:1939}  and is textbook material \cite{Luth:1992,Davison:1992}. In the present context, it is worthwhile pointing out that the characteristic exponential decay length $\kappa$ into the bulk depends on the energy difference $\Delta E$ between the surface state solution and the bulk states, with the order of magnitude given by $\kappa \approx \sqrt{2 m \Delta E / \hbar^2}$. For a mid-gap state in a topological insulator, this would be quite small, in the order of 1~nm but it increases as the state approaches the projected bulk states.  Finally, bulk states can be localized close to the surface due to quantum-confinement, either by band bending near the surface (or interface) of a semiconductor, as routinely realized in semiconductor devices but also found at pristine surfaces \cite{Colakerol:2006,Piper:2008,King:2008,King:2008c,King:2010}, or in thin metal films on semiconductors or insulators \cite{Chiang:2000,Milun:2002}. In the case of a semiconductor, the situation is schematically shown in Fig. \ref{fig:near_surface_states}(c). The band bending leads to a quantisation of the conduction band states near the surface. The wave function for the lowest quantum well state (QWS)  is schematically shown. On the mesoscopic scale, it is the solution of the Schr\"odinger equation obtained for a potential well of this shape. On a microscopic scale  the wave function retains its original Bloch character \cite{Bastard:1988,Chiang:2000,Milun:2002}. For our purpose, it is worth noticing that the quantum confinement turns the state from three-dimensional to two-dimensional in the sense that there is no dispersion in the direction perpendicular to the surface. The state is also confined to a region close to the surface but its decay away from the surface is not exponential and it typically penetrates much deeper into the material than a genuine surface state. Indeed, the physics of these quantum-confined states is very different to that of a surface states as they are directly derived from bulk states by merely restricting the number of allowed $k_z$ values. 

\begin{figure}[htbp]
		\includegraphics[width=6in]{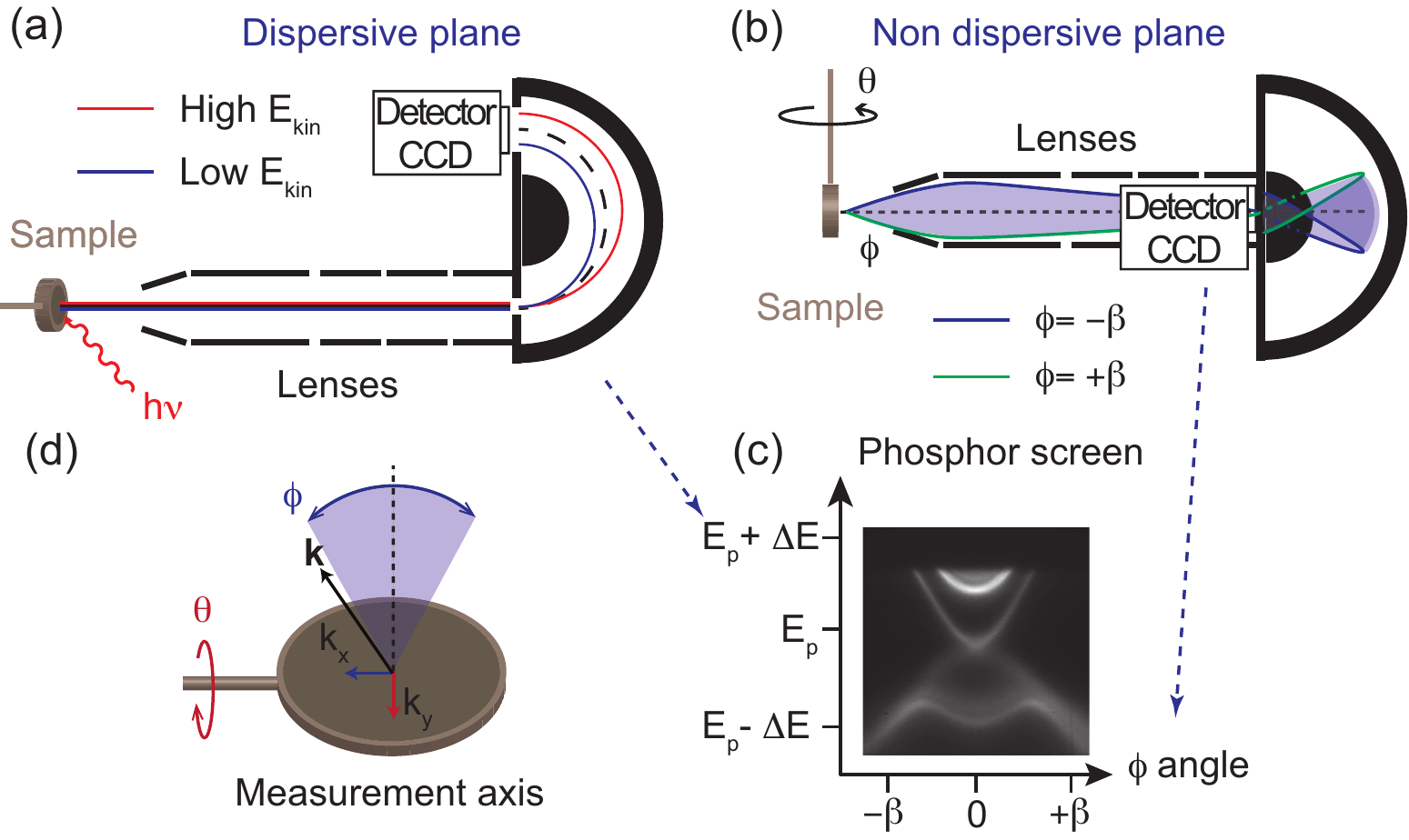}
		\caption{\footnotesize Scheme depicting the hemispherical analyser and the geometry of the ARPES experiment. (a) Cross section and definition of the dispersive plane of the analyser with the orbits of the higher (lower) kinetic energy electrons in red (blue). (b) Cross section of the non dispersive plane of the analyser with the orbits of the electrons emitted at an angle $+\beta$ ($-\beta$) in green (blue). (c) CCD picture of the acquired spectra on the detector. d) Scheme presenting the $\theta$ and $\phi$ angles used to map the photoemission intensity in the experiments performed.}
		\label{fig:kxky}
\end{figure}

When considering ARPES from the above states, it is useful to first establish the purely kinematic conditions for the state's observation. The introduction of the surface breaks the translational periodicity of the crystal in the $z$ direction and the wave vector in that direction $k_z$ is thus no longer well-defined. The 
components of the wave vector parallel to the surface ($k_\|$), on the other hand, are still well defined and must be  conserved in the photoemission process. They are thus obtained from the photoemission angle with respect to the normal of the sample and the kinetic energy of the photoelectron. 
In order to map the photoemission intensity on the hemisphere over the sample's surface, it is possible to implement an azimuthal scan (rotating the sample around the surface normal direction) for every polar angle considered. In today's ARPES setups, the electron analyzer often contains a two-dimensional detector that can simultaneously image the energy dispersion of the electrons and the angular dispersion in one direction (see Fig.  \ref{fig:kxky}). In this case, it is useful to consider the dispersion along two polar angles ($\theta$ and $\phi$), along orthogonal directions as presented in Fig. \ref{fig:kxky}(d). 
 In this configuration, the relation between the angles and $\mathbf{k}_\|$ is
 \begin{equation}
\mathbf{k}_\|= (\sin{(\phi)} \mathbf{\hat{x}}+\cos{(\phi)} \sin{(\theta)} \mathbf{\hat{y}})\cdot \sqrt{\frac{2 m_{e} E_{\text{kin}}}{\hbar^2}},
\end{equation}
where $E_{\text{kin}}$ is the kinetic energy of the photoemitted electron. 

\begin{figure}[htbp]
		\includegraphics[width=4in]{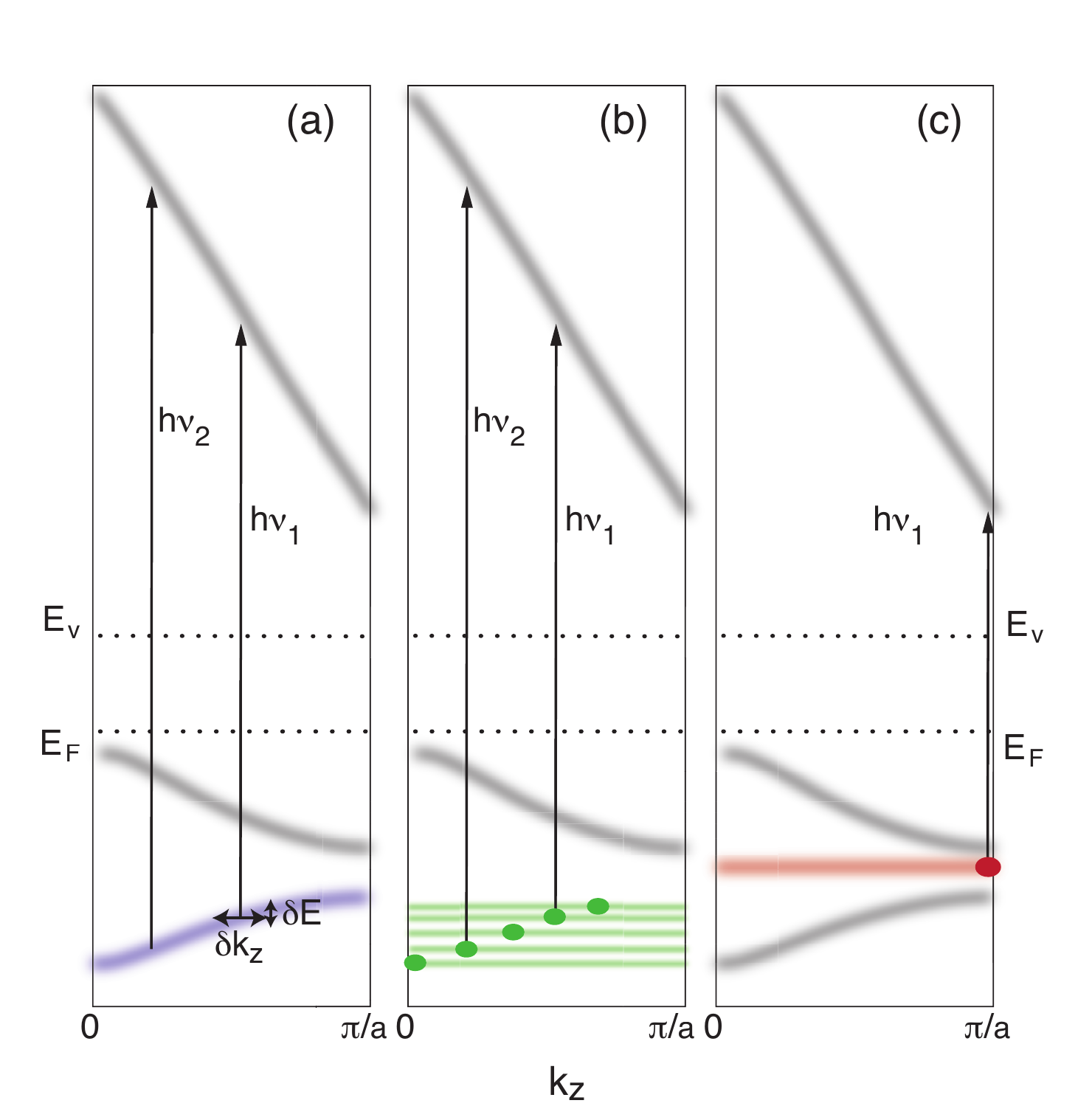}
		\caption{\footnotesize Schematic picture of the photoemission process from different types of electronic states. $E_F$ and $E_V$ are the Fermi and vacuum level, respectively. (a) Bulk states measured with different photon energies  appear at different binding energies in the spectrum, i.e. they show dispersion. The fact that $k_z$ is not well-defined leads to a broadening $\delta k_z$ that is then reflected in an energy broadening $\delta E$ of the observed peaks. (b) Quantum well states are derived from restricting the possible $k_z$ values in a structure (green markers, only shown for the lowest band) but the $k_z$ broadening (green lines) implies that they can be observed for a range of photon energies. As they do not show dispersion, this will not lead to an energy broadening. Similar arguments hold for surface states (c).  }
		\label{fig:photoemission_res}
\end{figure}

As pointed out above, the wave vector in the $z$ direction is not a good quantum number near the surface because the translational symmetry in this direction is broken. Even if we ignore this fundamental problem for the time being, it would not be possible to recover $k_z$ inside the sample from the value measured outside due to the potential change at the surface that leads to a refraction effect at the surface barrier \cite{Hufner:2003}. Despite these difficulties, it is often still possible to recover $k_z$ of the initial state from the photoemission spectra, provided that the dispersion of the final states is known. This is illustrated in Fig. \ref{fig:photoemission_res}(a). As the photon energy is changed, different initial state energies and $k_z$ values can be probed and the bulk state in question will appear as a peak at different binding energies in the spectrum, depending on the photon energy used.  The state's binding energy at high symmetry points ( $k_z= n\pi/a$ where $n=0, 1, 2...$ and $a$ is the one-dimensional lattice constant perpendicular to the surface) can easily be read from the spectra, but the detailed dispersion of the state can only be recovered if the final state dispersion is known. 

Even if this is not the case, it is often possible to obtain a good approximation of the entire dispersion by assuming final states of a simple form, for example free electron final states. In this case it is possible to calculate $k_z$ plus or minus a reciprocal lattice vector as a function of the emission angle $\theta$ and the inner potential $V_0$ using 
\begin{eqnarray}
k_z=\sqrt{^{2m_e} /_{\hbar^2} (V_0+E_{\text{kin}} \cos{(\theta)})}.
\label{eq:kz}
\end{eqnarray}
The inner potential $V_0$ can be determined iteratively by requiring the resulting $k_z$ to be consistent with the binding energy extrema at symmetry points. 

The problem that $k_z$ is not well-defined anymore cannot be neglected. In particular, the finite escape depth of the photoelectron leads to a $k_z$ broadening $\delta k_z$. As the state disperses in the $k_z$ direction, this also leads to an energy broadening $\delta E$, as illustrated in Fig. \ref{fig:photoemission_res}(a).

This problem is absent for photoemission from surface states and quantum-confined states, shown in Fig. \ref{fig:photoemission_res}(b) and (c), respectively (again, the quantum confinement has only been introduced for the lowest band). Strictly spoken, neither the surface state nor the quantum confined state have a well-defined $k_z$ because they are localized in the $z$ direction. This is usually a much more severe restriction for the more strongly confined surface states than for the deeply penetrating QWS. In neither case does the wave function loose its periodicity perpendicular to the surface completely and the Bloch wave character in the $z$ direction is partly retained. Thus, the markers in the figure symbolize the $k_z$ value the state is derived from and the horizontal lines symbolize that $k_z$ is not defined anymore. In any case, the absence of a dispersion with $k_z$ implies directly that a $k_z$ broadening does not result in an energy broadening in the photoemission spectra, unlike the case for the bulk states. 
This implies that it is often easier to determine a bulk band structure in the presence of quantum confinement than in its absence \cite{Chiang:2000}. The fact that $k_z$ is ill-defined also implies that surface states and QWS can be observed not only for one photon energy but for a broad range of photon energies.

With this we come to the dynamic properties of the photoemission from bulk states, surface states and QWS, and we consider the photoemission intensity we can expect to measure. Even if we assume perfect angular and energy resolution in the experiment, the complete expression is rather complicated \cite{Matzdorf:1998}. One obtains:
  \begin{equation}
I(E_{kin},\mathbf{k}) \propto |M_{fi}(\mathbf{k}_f,\mathbf{k}_i)|^2 f(h\nu-E_{kin}-\Phi,T) \int
\mathcal{A} (h\nu-E_{kin}-\Phi,\mathbf{k})
\mathcal{L}(k_{z},k_{z}^0)dk_{z},
 \label{equ:p2}
 \end{equation}
where $\mathbf{k}_i$ and $\mathbf{k}_f$ are the three-dimensional wave vectors for the initial and final state in the solid,  $\mathbf{k}$ is the wave vector for the free electron outside the solid, $h\nu$ is the photon energy, $f$ is the Fermi distribution, $\Phi$ is the work function, $ \mathcal{A}$ is the hole spectral function and $
\mathcal{L}$ is a Lorentzian distribution to account for the
real-space damping of the outgoing electron wave, with $k_{z}^0$ being the
perpendicular wave vector for the un-damped final state wave.   $M_{fi}$ is the matrix element for the photoemission process. The integration is over all possible perpendicular wave
vector components $k_{z}$.  For QWS or surface states, this integration becomes irrelevant. 

The quantity of most interest is often the spectral function $\mathcal{A}$. In simple terms, this is a measure of finding an electron with a certain energy and crystal momentum, i.e. an image of the band structure, including lifetime broadening effects. For two-dimensional systems such as surface states and QWS, the photoemission intensity is directly proportional to $ \mathcal{A}$, as no final state broadening has to be considered. For a detailed further analysis it is only necessary to assume that the matrix element does not depend strongly on the binding energy or wave vector. 

In the present context, the matrix element depends rather strongly on the photon energy  and this dependence contains valuable information on the nature of the observed states. This is evident when considering Figs. \ref {fig:near_surface_states} and \ref{fig:photoemission_res}. For bulk states, no significant change in the photoemission intensity is expected as the photon energy is varied (assuming a gap-free final state). A different photon energy will merely lead to emission from a bulk state around a different $k_z$ (Fig. \ref{fig:photoemission_res}(a)). For  surface states and QWS (Fig. \ref{fig:photoemission_res} (b) and (c)) this is not so: both lack a well-defined $k_z$ but the wave functions still have a remaining Bloch character perpendicular to the surface. For the QWS, this Bloch character is very similar to the parent bulk states and for the surface state it is similar to the bands near the BZ boundary at $k_z=\pi/a$ (in this particular case). Thus, a resonant enhancement in the photoemission intensity of such states is expected for photon energies that would have lead to emission from the parent bulk states. This is indeed observed in many cases \cite{Louie:1980,Hofmann:2002} and it will be used in the following discussion. 

\section{Methods}


Samples grown from stoichiometric mixutures are known to be highly $n$-doped due to presence of defects in the bulk and mainly charged Se vacancies \cite{Navratil:2004}.  For this reason two different \bise~crystals were used during the experiments. The first batch of crystals, named as intrinsic, was obtained from a Se-rich mixture of 5N purity  elements (Bi : Se = 2 : 3.3)  melted at 860$^{\circ}$C for 24 hours in an evacuated quartz ampoule, cooled down to 650$^{\circ}$C at a rate of 2.5$^{\circ}$C/h, and then annealed at 650$^{\circ}$C for 7 days. A second batch, named as Ca-doped, was obtained by a mixture of Bi (5N purity) , Se (5N purity) and Ca (99.5\% purity) with ratio Bi : Se : Ca = 1.996 : 3 : 0.004 melted at 860$^{\circ}$C for 24 hours in an evacuated quartz ampoule. Once cooled down from 860$^{\circ}$C to 750$^{\circ}$C at a rate of 50$^{\circ}$C/h and then from 750$^{\circ}$C to 600$^{\circ}$C at a rate of 2~$^{\circ}$C/h, it was annealed at 600$^{\circ}$C for 7 days. The Ca doping is able to recover the insulating state and the long cooling time produces large crystals. The crystal structure and quality was determined using x-ray diffraction at room temperature. 

ARPES spectra were acquired at the SGM III undulator beamline at the synchrotron light source ASTRID \cite{Hoffmann:2004} which is equipped with a SPECS Phoibos 150 spectrometer. The  combined energy resolution (photons and electrons) in this work was better than 15~meV and the angular resolution was better than 0.13$^\circ$. 
After cleaving the sample \emph{in situ} and at room temperature (pressure better than $1\times 10^{-8}$ mbar), it was immediately transferred  to the measurement position at a pressure better than $4\times 10^{-10}$ mbar and cooled to a temperature of 60~K.  

Simple model calculations were used to determine the energy positions of the QWS in the near-surface  potential well that is caused by the downward bending of the conduction band.  To this end, the potential near the surface was simulated  using a  Schottky model. A constant charge density $\rho$ was assumed to be present between the surface ($z=0$) and a certain depth ($z=\Delta$). From this, the electrostatic potential was calculated via the Poisson equation. We then numerically solved the  Schr\"odinger equation in this potential and obtain the allowed eigentstates and their energies. 
The model has two parameters, $\rho$ and the width of the charge layer $\Delta$. These can be determined by two conditions. The first is that the potential must reproduce the experimentally observed band bending.  The experimental value can be inferred from the position of any sharp state in the spectrum. Either a valence state or a core level can be used \cite{Himpsel:1983b}. We have used the topological surface state for this purpose (see below). The second condition is that the solution of the 
Schr\"odinger equation must reproduce the number and approximate energies of the experimentally observed occupied QWS. The solution of the Schr\"odinger equation also yields the envelope wave functions as shown in Fig. \ref{fig:near_surface_states}(c). As usual with this type of calculation, the Bloch character of the states  is taken into account by using the correct effective mass of the electrons in the $z$ direction. Here this is set to 0.24~$m_0$ \cite{Bianchi:2010b}. {Note that this simple Schottky model is quite different from a previously used model \cite{King:2008,Bianchi:2010b} that takes nonparabolic bands into account via the  $\mathbf{k}  \cdot \mathbf{p}$ approximation and calculates the quantum well states via a coupled solution of the Schr\"odinger and Poisson equations \cite{King:2008}. The purpose of the calculations here is not a quantitative agreement between experiment and model or an extraction of parameters like the near-surface charge density. It is merely to have a fast calculation that can produce quantum well states for different degrees of band bending and this can be handled satisfactory by this and even simpler models, such as a triangular quantum well \cite{Benia:2011}.  }


\section{Results and discussion}

\begin{figure}[htbp]
		\includegraphics[width=5in]{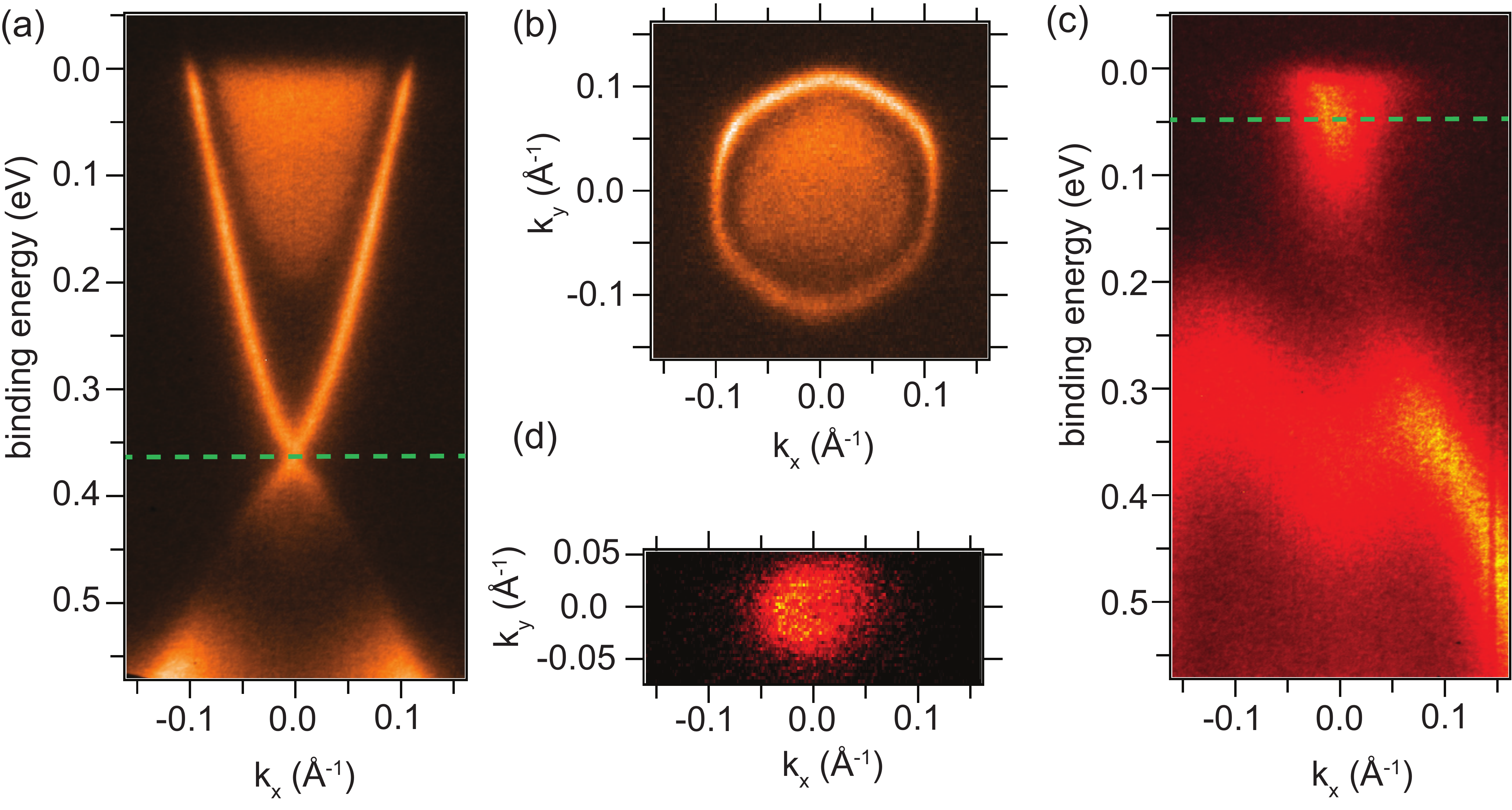}
		\caption{\footnotesize ARPES spectra for the pristine surface of \bise. High photoemission intensity is displayed in bright. (a) Energy dispersion in the $\bar{\text{K}}\bar{\Gamma}\bar{\text{K}}$ direction of the SBZ and (b) Fermi surface for the stoichiometric \bise~sample. (c) and (d) Energy dispersion and Fermi surface for the Ca-doped sample, respectively. The Dirac point is at a binding energy of $\approx 50$~meV.}
		\label{fig:fresh}
\end{figure}

We will start by discussing the  pristine surface of intrinsic and Ca-doped  \bise~crystals. 
Fig. \ref{fig:fresh}(a) and (b)  show  ARPES spectra taken from the two batches and in the simple picture described above, these can be viewed as cuts through the spectral function along a certain direction in the SBZ. Fig. \ref{fig:fresh}(a) shows the the dispersion of the states, i.e. the photoemission intensity as a function of $k$ in the $\bar{K}\bar{\Gamma}\bar{K}$~direction and binding energy, whereas  Fig. \ref{fig:fresh}(b) shows the photoemission intensity at the Fermi level as a function of $\mathbf{k}_\|$. Both spectra are cuts through a three-dimensional data set in which the photoemission intensity is collected as a function of binding energy and the two components of $\mathbf{k}_\|$.  The distinct V-shaped feature in Fig. \ref{fig:fresh}(a) stems from the topological surface state, forming a conical dispersion with its  Dirac point at a binding energy of 360~meV. At the Fermi level, i.e. far away from the Dirac point, the dispersion deviates from a simple cone and the constant energy surfaces are hexagonal (Fig. \ref{fig:fresh} (b)) \cite{Kuroda:2010,Bianchi:2010b}. The anisotropy of the Fermi velocity is known and explained for the analogous Bi$_2$Te$_3$ case where the deviation from a perfect Dirac cone is more evident \cite{Chen:2009,Fu:2009}. The features at the highest binding energies and the region of high intensity inside the Dirac cone are due to emission from the bulk valence and conduction bands, respectively. Despite the Se enriched mixture used in the growth, these samples are still degenerately $n$-doped. We also see that the spectral features linked to the topological state, a genuine surface state, are considerably sharper than those of the other states. This is due to the $k_z$-smearing explained in Fig. \ref{fig:photoemission_res}.

The Ca doping is able to counteract the natural doping, resulting in a shift of the Dirac point from a binding energy of 365~meV  to 47~meV as shown in Fig. \ref{fig:fresh} (c) and (d). In this case, the topological state appears broader, consistent with the higher degree of disorder due to the random distribution of dopants in substitutional Bi sites \cite{Hor:2009}. Such disorder is relevant even for surface states, as it induces an additional smearing in $k_\|$. 

\begin{figure}[htbp]
		\includegraphics[width=4.5in]{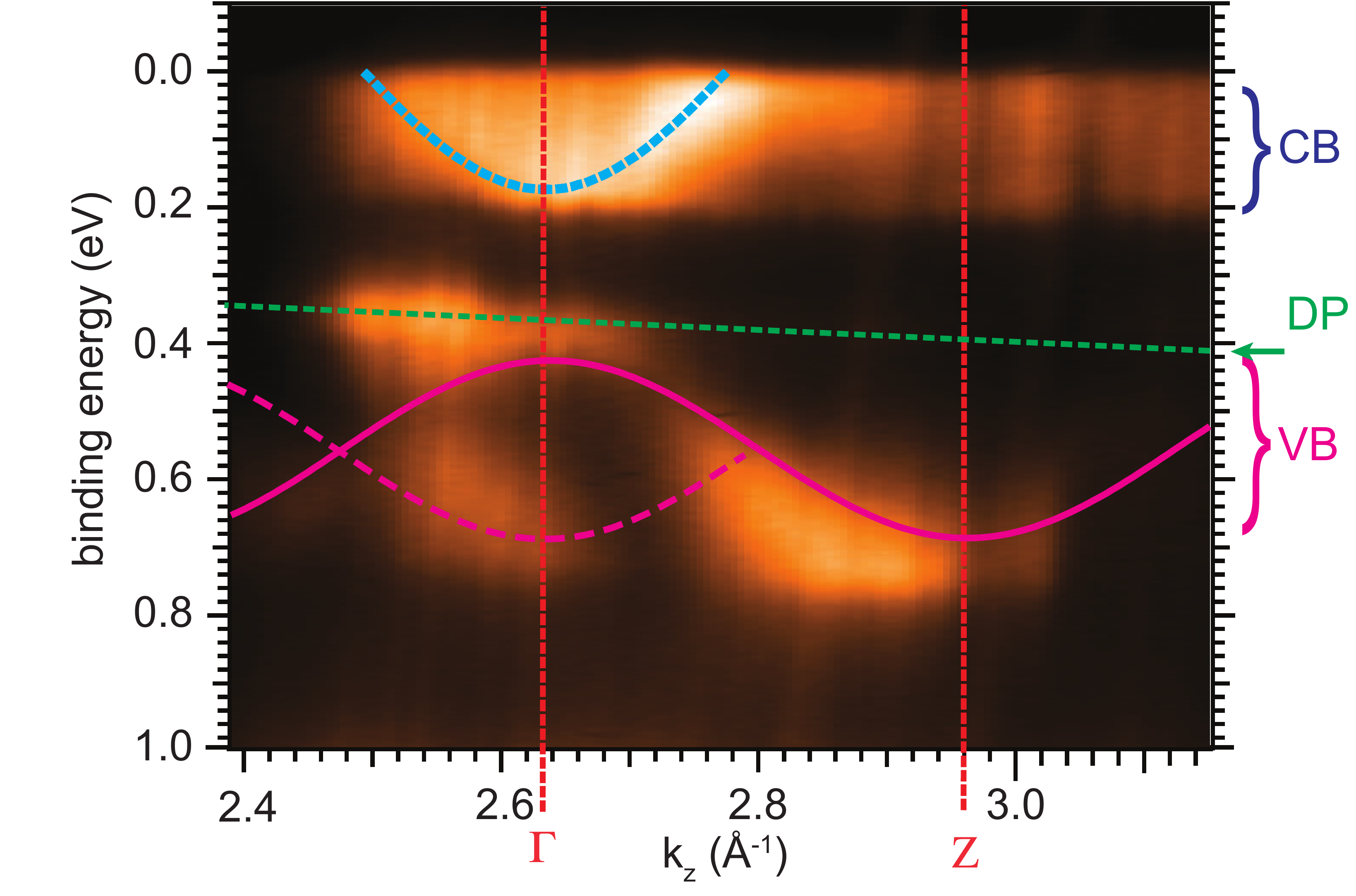}
		\caption{\footnotesize Photon energy scan on the pristine surface of intrinsic \bise~ crystal illustrating the dispersion of the states at normal emission as a function of $k_z$. The data shown are a subset of a larger photon energy scan between $h\nu = 14$ eV and $h\nu = 32$ eV. The drift of the Dirac point with photon energy is due to the ageing effect that occurs during the scan. The CB and VB  (highlighted with blue and magenta lines as a guide to the eye) disperse, revealing the bulk $\Gamma$ and $Z$ points. The dashed magenta line is a shifted replica of the valence band dispersion caused by a surface umklapp process.}
		\label{fig:FreshEscans}
\end{figure}

The dispersion of the bulk states in the direction perpendicular to the surface can be inferred from data taken as a function of the photon energy, such that different values of $k_z$ are probed (see Fig. \ref{fig:photoemission_res}(a)). 
Fig. \ref{fig:FreshEscans} shows the result of such a photon energy scan, presenting the photoemission intensity at normal emission acquired for the intrinsic sample. Data were taken as a function of photon energy, but the horizontal axis has already been transformed to $k_z$ values instead of photon energies, using (\ref{eq:kz}).  For this transformation, an inner potential of $V_0$~=11.8 eV was assumed \cite{Xia:2009,Bianchi:2010b}. 

  The dispersing features stem from the conduction band (CB, blue) and valence band (VB, magenta). 
The bottom of the conduction band  appears together with the top of the VB at $k_z\approx 2.63 $\iA (h$\nu \approx 19.2$  eV), corresponding to the bulk $\Gamma$ point.  The bottom of the VB is reached at $k_z\approx 2.96$\iA (h$\nu \approx$ 26.6 eV) and corresponds to the bulk $Z$ point. 
This corresponds to a $\Gamma Z$ distance of 0.329\AA$^{-1}$ and is consistent with the literature \cite{Wyckoff:1963}. 
The green line in the scan corresponds to the Dirac point of the surface state. The position of the Dirac point cannot be easily read from this scan but it is inferred from the dispersion of the topological state in images such as Fig. \ref{fig:fresh} (a). In fact, the data set of Fig. \ref{fig:FreshEscans} was constructed by taking the centre intensity of many of such images taken at different photon energies. Note that the position of the Dirac point does not appear at a strictly constant energy, even though this should be expected for a two-dimensional state. The reason for this is the ageing effect with the time-dependent downward band bending. Taking a data set large enough to construct Fig. \ref{fig:FreshEscans} takes approximately 30 min, and this is long enough for the ageing effect to be noticeable. 

In addition to the expected valence band dispersion marked by the magenta line, a second identical but weaker dispersion is found (dashed magenta line). This is shifted by half a reciprocal lattice vector in the $k_z$ direction, i.e. it has its highest binding energy at $\Gamma$ instead of $Z$. The presence of this band is attributed to an umklapp process involving a surface reciprocal lattice vector. Such processes are well known for similar BZ shapes,  where the $k_z$ value of the $\Gamma$ point in the first Brillouin zone is the same as for the zone boundary in the neighbouring zone \cite{Hofmann:2002}.  

\begin{figure}[htbp]
		\includegraphics[width=6in]{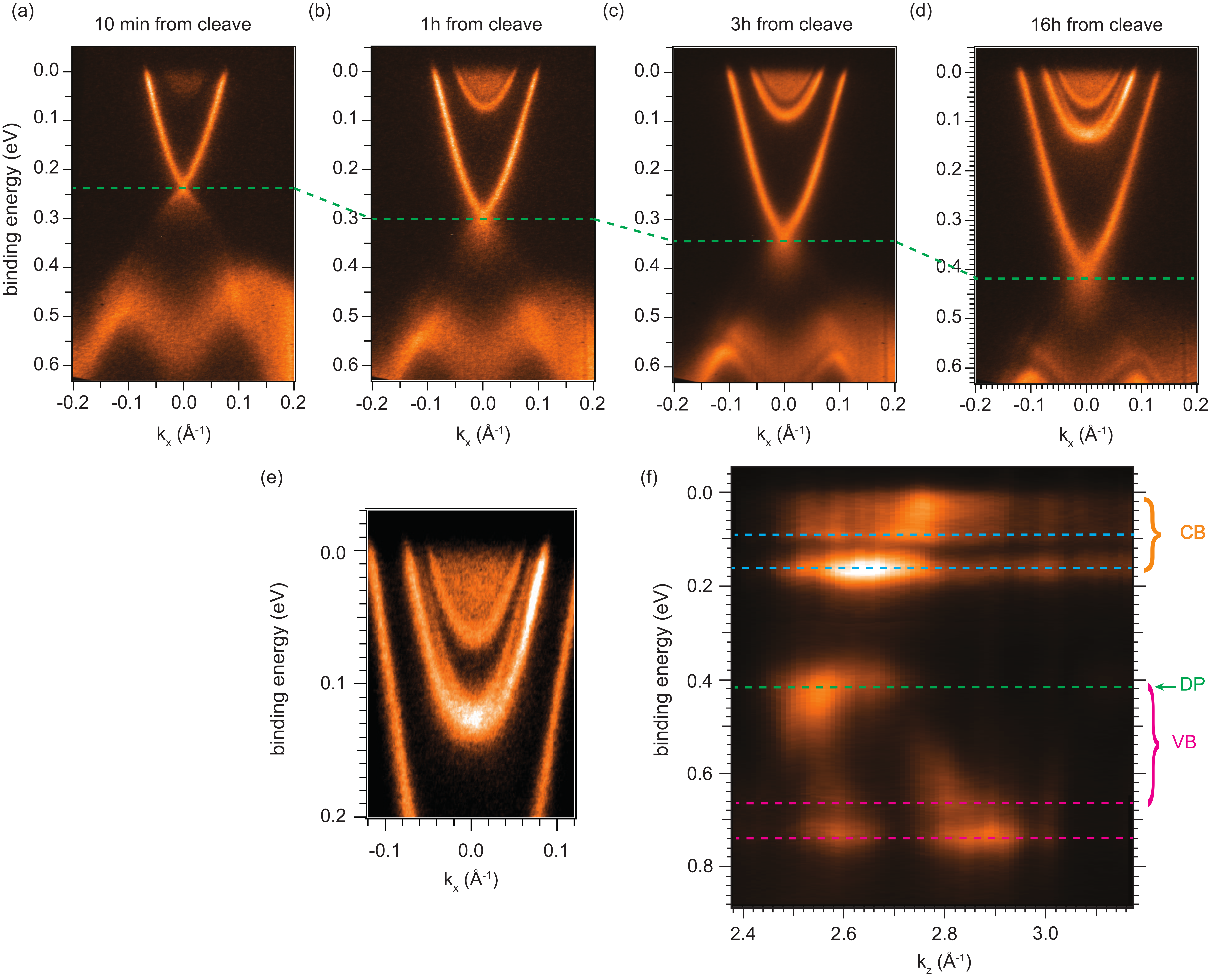}
		\caption{\footnotesize ARPES spectra acquired at (a) 10 min, (b) 1 h, (c) 3 h and (d) 16 h after cleave showing photoemission intensity as a function of binding energy and $k_\|$ ($hv=16$ eV). Additional states in the CB are clearly visible. The band bending is indicated by the shift of the Dirac points towards higher binding energy. (e) Magnification of the Rashba split quantum well state of figure (d). (f) Photoemission intensity at normal emission as a function of binding energy and $k_z$ acquired with a photon energy scan. The data shown in (f) are a subset of a larger energy scan between $h\nu = 14$ eV and $h\nu = 32$ eV.}
		\label{fig:timeDep}
\end{figure}

Fig. \ref{fig:timeDep} provides a more detailed picture of the ageing effect. As soon as one hour after cleaving the sample, a clear shift of the Dirac point is observable together with the appearance of additional states: a parabolic state at the rim of the CB and a sharp M-shaped state in the  VB region. As time elapses the band bending increases and the number of additional states increases. Eventually, two distinct states are found the CB region, with the lower state showing a clear Rashba splitting \cite{King:2011}. 
The expected spin-texture of these states has been confirmed by spin-resolved ARPES \cite{King:2011}.
The ageing also leads to a small broadening of the topological state, indicating surface disorder  \cite{Hatch:2011}.
It should be stressed that the drift of the photoemission features towards higher binding energies is a well known phenomenon for semiconductors, indicating a downward shift of the electronic bands near the surface \cite{Noguchi:1991,Watanabe:1997,Deng:2000,Lowe:2002}

The states in the conduction band have been interpreted as QWS caused by the strong band bending near the surface. The situation for the lower conduction band QWS corresponds to that in Fig. \ref{fig:near_surface_states}(c) and its energy position can indeed be reconciled with the observed amount of band bending, using either a simple Schottky model (see below) or a more sophisticated solution of the coupled Poisson and Schr\"odinger equations \cite{King:2008,Bianchi:2010b}. 
The two-dimensional character of the states can be tested directly by a photon energy scan. The result is given in Fig. \ref{fig:timeDep}(f) and should be compared to Fig. \ref{fig:FreshEscans}. The difference in the conduction band region is clear: in the case of weak band bending, the dispersion of the entire conduction band is clearly seen, but for stronger band bending only two non-dispersive features are observed at similar energies. This is a clear experimental indication for the formation of QWS. It is also seen that the intensity of the non-dispersive feature is resonantly enhanced at the $\Gamma$ point, i.e. at the energy where previously the non-confined state was observed. 

In the following we will argue that the band bending is induced by the adsorption of impurities. Indications of this are the stronger tendency towards ageing at lower temperatures and higher background pressures and that the band bending can be partly reversed by annealing the surface to higher temperatures \cite{King:2011}. We will demonstrate how the same ageing effect can be obtained in a controllable and tunable way by depositing known species on the surface. 

To test which species can cause the band bending, the clean surfaces have been exposed to a controlled dose of carbon monoxide  \cite{Bianchi:2011}, as CO represents one of the common rest gas molecules present in an ultra-high vacuum recipient. Similar tests have been performed with CO$_2$ and N$_2$ but this did not reveal an appreciable acceleration of the normal ageing effect (data not shown), apart from an increased broadening and a signal degradation.

\begin{figure}[htbp]
		\includegraphics[width=6in]{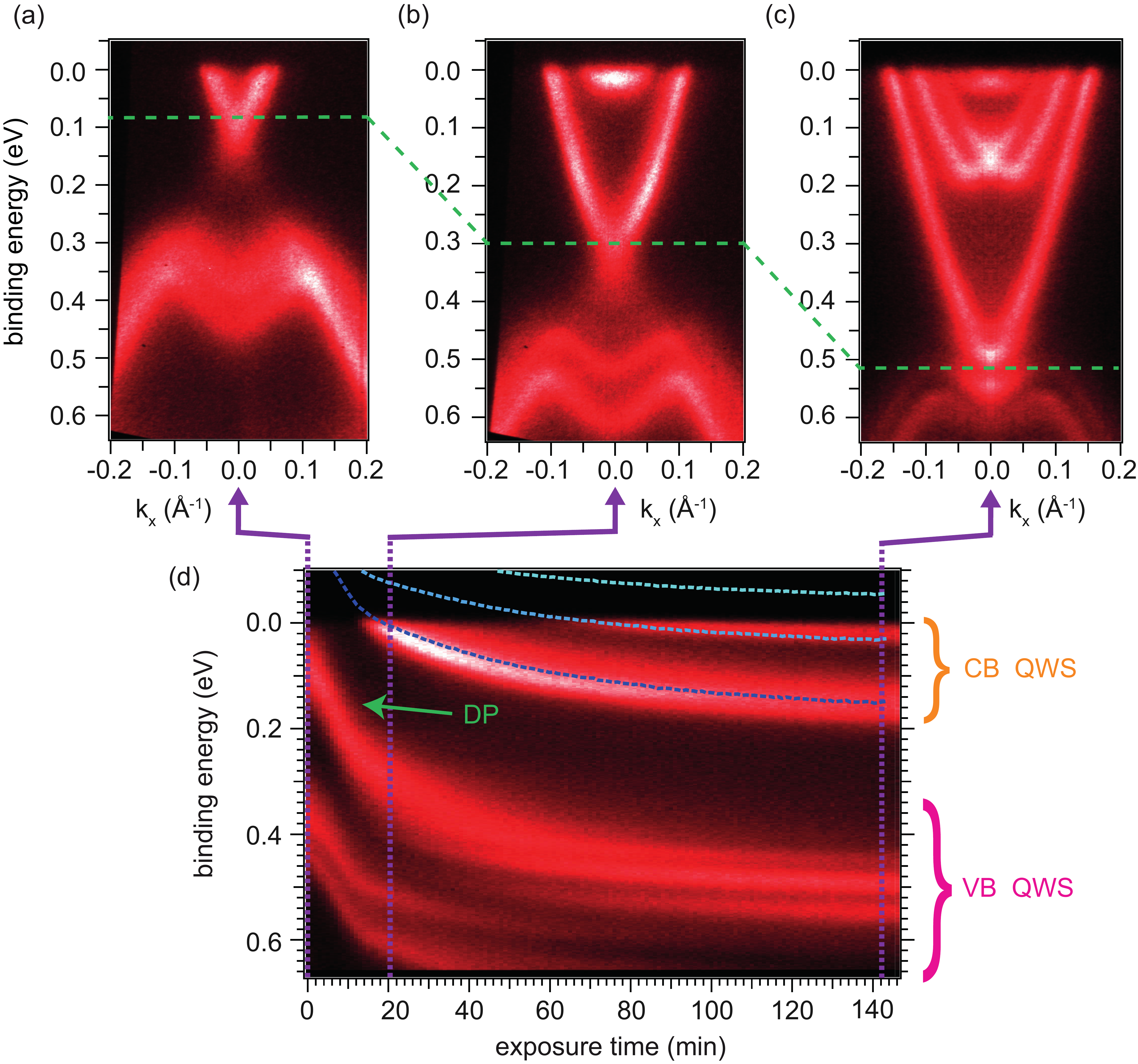}
		\caption{\footnotesize ARPES spectra acquired during the controlled dose of CO  on Ca-doped \bise~($h\nu= 16 $ eV). The acquisition and dosing was started about 1~h after the cleave. (a)-(c) Photoemission intensity as a function of binding energy and $k_\|$ at different exposure times. (d) Photoemission intensity at normal emission as a function of binding energy and exposure time. The states in the marked regions are the quantum well states (CB QWS and VB QWS). The Dirac point is indicated by the green arrow while the blue, azure and cyan dashed lines show the energies of the calculated QWS.}
		\label{fig:COtime}
\end{figure}

Fig. \ref{fig:COtime} is analogous to the time dependence shown in Fig. \ref{fig:timeDep}, but here spectra were acquired at a partial pressure of $8 \times 10^{-9}$ mbar of CO on a  Ca-doped sample that was kept at a temperature of 65 K. Acquisition was started 1 hour after the cleave.  The development of the band bending is faster and more pronounced than before. Fig. \ref{fig:COtime}(a)-(c) show the dispersion at different times. These spectra are merely a small subset of a large series of images. A cut through the centre of all images (i.e. the energy and time-dependent photoemission intensity at normal emission) is shown in Fig. \ref{fig:COtime}(d). This illustrates the position of the states as the band bending increases. In contrast to Fig. \ref{fig:timeDep}, states of the non-confined conduction band are never observed. This is ascribed to the fact that we start out with a Ca-doped sample and the conduction band only becomes visible when the band bending is already considerable. For a sufficiently large dose of CO, after circa 20 minutes of exposure, the band bending is strong enough for the lowest CB QWS to move below the Fermi level and thus become observable. In addition to the time of exposure, the band bending values are given on the horizontal axis. The values have been extracted from the position of the topological state in images like Fig. \ref{fig:COtime} for a non-zero value of $k_x$, where the state is always clearly observable, even for strong band bending.  
Saturation is reached when the band bending is stabilized and the the binding energies of the QWS have reached a steady value.
The saturation coverage of CO has not been determined but it is expected to be $\lesssim 1$ monolayer. 
Multilayer formation has not been observed and it is not expected on inert surfaces at these temperatures and pressures. 

Fig. \ref{fig:COtime}(d) also shows the results of the Schottky model simulation. The energies of the calculated QWS are plotted as blue, azure and cyan dashed lines on top of the data. Clearly, the agreement between the simple model and the experiment is very good. As discussed previously, the model has two free parameters, the charge density $\rho$ and the depth below the surface $\Delta$. These parameters can be adjusted to achieve a good fit of the QWS energies but here they are only adjusted for one value of the coverage and then kept fixed for all others. 

\begin{figure}[htbp]
		\includegraphics[width=6in]{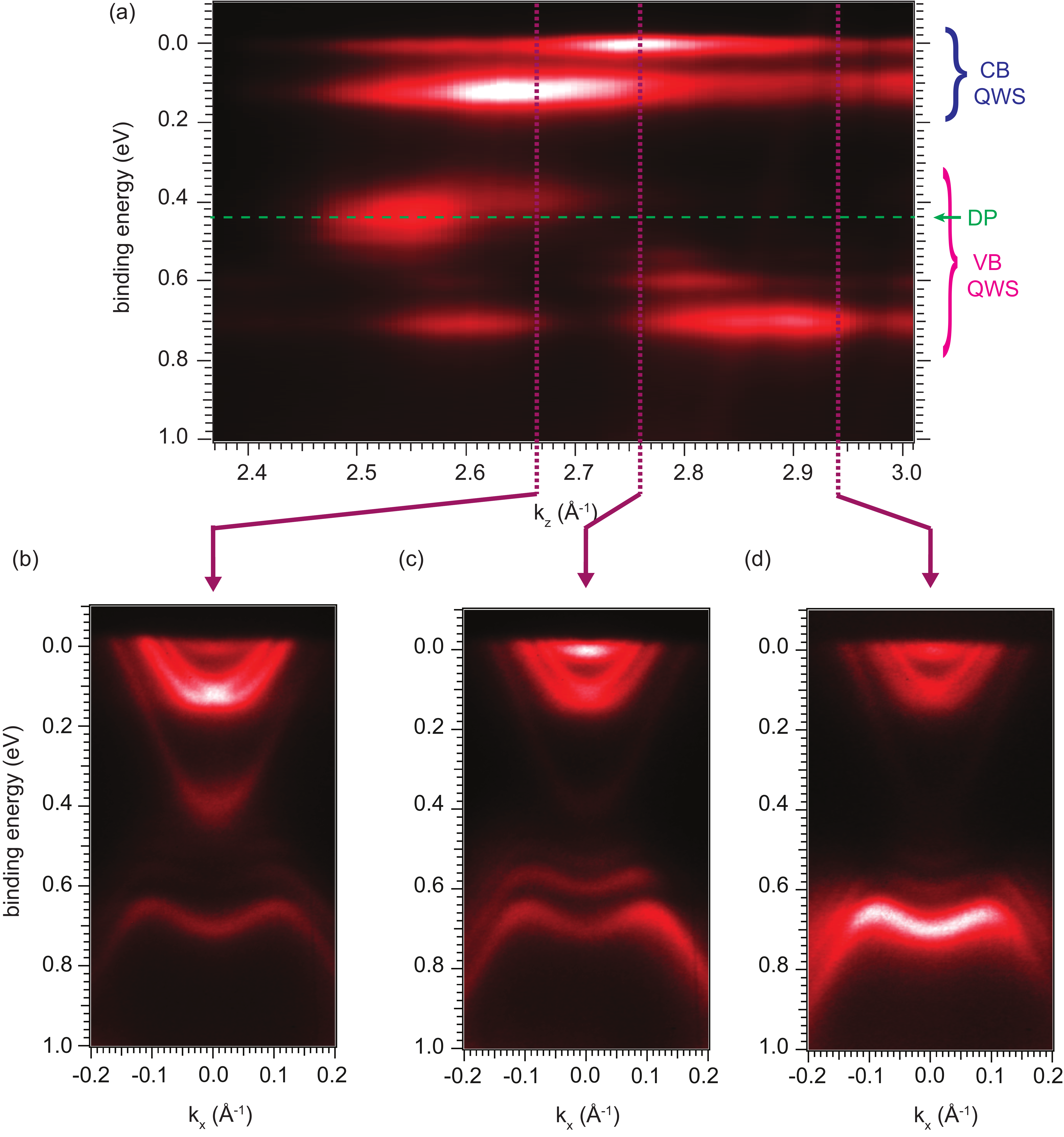}
		\caption{\footnotesize Photon energy scan after exposing the Ca doped samples to carbon monoxide. (a) Photoemission intensity at normal emission in function of binding energy and $k_z$. (b)-(d) Photoemission intensity at selected $k_z$ values depicting the energy dispersion relation} 
		\label{fig:COEscan}
\end{figure}

Again, a  photon energy scan confirms the two-dimensional character of the of the conduction band QWS (Fig. \ref{fig:COEscan}). It also shows that the VB states are now quantised, too, with the previously continuous valence band now being replaced by distinct features. These are also visible Fig. \ref{fig:COtime}, and have a characteristic  M-like shape. This simultaneous quantisation of both valence band and conduction band states is rather unusual and related to the large projected band gap under the top part of the VB shown in  Fig. \ref{fig:BzCrystal}(c). It is explained in Ref. \cite{Bianchi:2011} and shall not be discussed in much further detail here. 
However, it is interesting to note the enhancement of the M-shaped states for particular photon energies. This happens at the same energies as the emission from the corresponding non-quantised valence band states through direct photoemission ( in the region of $h\nu \approx 25.3$ eV) and via the umklapp process with a surface reciprocal lattice vector ($h\nu \approx 19$  eV). This behaviour is consistent with what would be expected for quantum confined states in the VB and with the simple picture given in Fig. \ref{fig:photoemission_res}.

\begin{figure}[htbp]
		\includegraphics[width=6in]{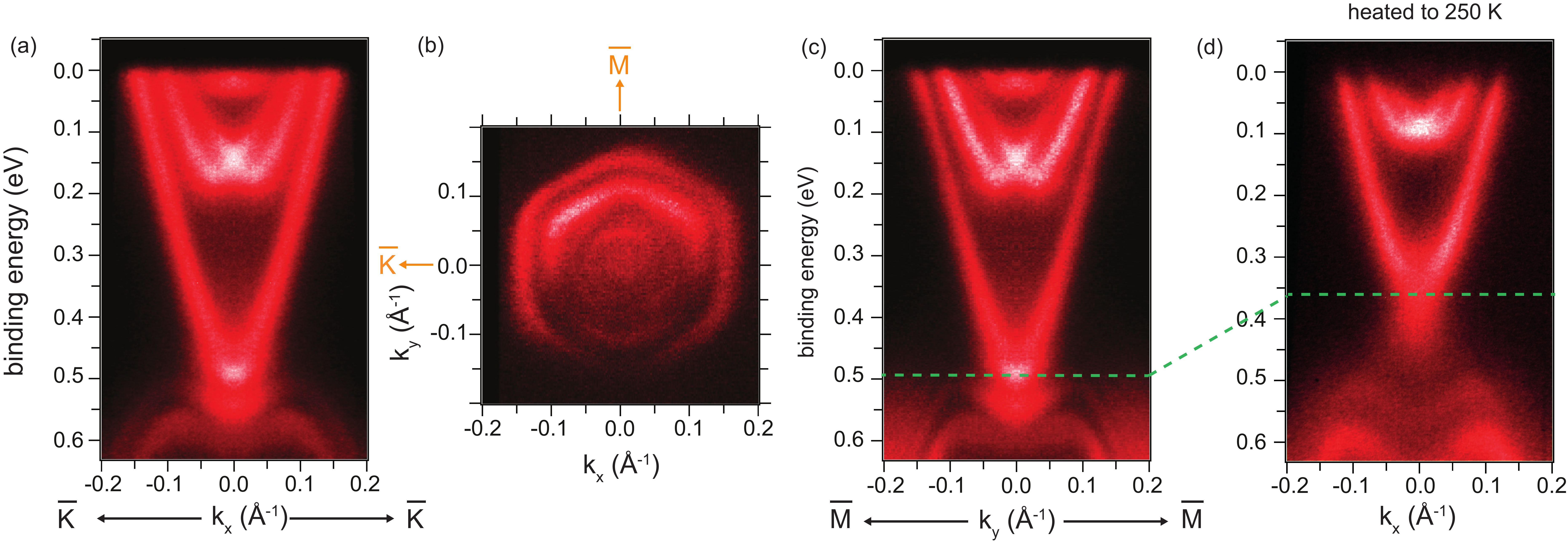}
		\caption{\footnotesize Dispersion and Fermi surface map after CO dosing ($h\nu=16$ eV) on Ca doped samples. (a) and (c) Photoemission intensity in function of binding energy and $k_\|$ in the $\bar{\text{K}}\bar{\Gamma}\bar{\text{K}}$ and $\bar{\text{M}}\bar{\Gamma}\bar{\text{M}}$ directions, respectively. (b) Fermi surface map around $\bar{\Gamma}$ showing the hexagonal warping of the QWS that diminishes as the distance from $\bar{\Gamma}$ decreases. (d) Spectrum taken at 250 K after CO dosing showing the partial reversibility of the band bending.}
		\label{fig:COFS}
\end{figure}

{
Interestingly, in both this work, and for Bi$_2$Te$_3$ \cite{Chen:2012}, the QWS are not found to be Rashba-split, in contrast to the QWS in the conduction band. The reason for this is likely  the different microscopic wave functions of the VB and CB states. It is well-known that the Rashba splitting of extended states is largely an atomic effect and a detailed understanding of the size \cite{Bihlmayer:2007} and even the sign \cite{Bentmann:2011} of the splitting requires a knowledge of the atomic wave function near the nuclei. Indeed, the different wave function character is already evident from the state's dispersion near $\bar{\Gamma}$ where the VB states are much flatter and a small Rashba splitting, being an offset in $k$, would be hard to detect.} 

The Fermi surface map presented in Fig. \ref{fig:COFS} demonstrates that the QWS and the topological state undergo an similar hexagonal warping. 
Both show weaker warping occurring at smaller momentum, as described by the model in \cite{Fu:2009}. 
While the inner branch of the QWS leads to a circular Fermi contour, the outer one appears hexagonal.

Recently, it has been shown that the dosing of H$_2$O leads to similar effects as observed for CO \cite{Benia:2011}. The band bending is similar but its origin has been ascribed to a chemical reaction occurring at  the surface that extracts Se atoms, leaving positively charged vacancies. In the case of CO adsorption, such a reaction appears unlikely. CO is a strongly bonded molecule (triple-bond), and it can only be dissociated by adsorption on reactive transition metal surfaces. {The adsorption of CO as a molecular species  is consistent with the behaviour of the surface electronic structure upon mild annealing of the sample (see Fig. \ref{fig:COFS} (d)). When we heat the sample to 250~K, the band bending and Rashba-splitting is reduced, consistent with a partial desorption of CO. Note that the thermal desorption of CO from Bi$_2$Se$_3$ has not yet been studied in any detail, and it is not clear if the
adsorption can be completely reversed by thermal desorption. A recent theoretical study of molecular adsorption of other species on Bi$_2$Se$_3$ suggests that this might not be the case due to preferential and strong bonding to defect sites on the surface  \cite{Koleini:ARXIV}.}

We now return to possible alternative explanations of the phenomena discussed here. Different interpretations have been proposed, in particular to explain the Rashba-split states in the CB region but also for the M-shaped states in the VB region. Wray \emph{et al.} show that states of similar appearance can be found in a slab calculation of the electronic structure of \bise~when the energies of the states near the surface are rigidly shifted \cite{Wray:2011}. Such a procedure can be viewed as a crude implementation of a band bending and is thus not dissimilar to the picture given here. {In fact, it is otherwise very difficult to accurately account for the band bending in a supercell-type density functional theory calculation. The reason is that the the slab used in such a calculation usually contains a limited number of layers, typically between 12 quintuple layers (114 \AA)~\cite{Xia:2009} and 25 quintuple layers (240 \AA)~\cite{Zhang:2010}. Band bending effects that extend to $\approx 300$~\AA~into the sample \cite{Bianchi:2010b}, are thus difficult to describe accurately with the limited slab size.}

An alternative scenario is discussed in Refs. \cite{Eremeev:2011b,Menshchikova:2011,Ye:2011}. There it is argued that spectroscopic changes similar to those reported here could be caused by an increase in the van der Waals gap(s) between the near-surface quintuple layers. This increase of van der Waals gap sizes could be caused by the intercalation of adsorbed atoms or molecules. Strongly increased distances between quintuple layers (between 10\% and 40\%) are predicted to lead to  two major changes in the electronic structure: the appearance of Rashba-split states that are split-off the conduction band minimum and M-shaped states in the projected band gap below the valence band maximum (see Fig. \ref{fig:BzCrystal}(c)).
It is indeed  plausible that at least alkali atoms can be intercalated into the van der Waals gap in \bise, as this is a well-known phenomenon for bulk crystals \cite{Paraskevopoulos:1988}, but for molecules such as CO intercalation appears unlikely. On the other hand, this proposed mechanism fails to explain a key experimental observation;  a calculation with an increased interlayer spacing predicts a single M-shaped state in the valence band region whereas several such states are observed experimentally. 
{This shortcoming can probably be improved  by allowing for several van der Waals interlayer spacings to relax \cite{Ye:2011}. A more important issue than the number the theoretically predicted M-shaped states is that their energy position of  does not agree with the experimental result. In the calculation, the M-shaped states are found in the projected band gap below the top of the valence band at $\bar{\Gamma}$. These states are  thus genuine new surface states. In the experiment, on the other hand, the M-shaped states are found within the region of the projected valence band, with the possible exception of the state at highest binding energy. This is directly evident from comparing the data shown in Figs. \ref{fig:timeDep} and \ref{fig:COEscan}.} These states, therefore, cannot be surface states, but must be QWS derived from the bulk states. Another interesting point is that the calculated surface electronic structure is only significantly modified for rather large changes in the van der Waals interlayer spacing, on the order of 10~\% or more. Experimentally, the intercalation-induced changes are often found to be rather small. The intercalation of substantial amounts of copper into bulk \bise, for example, leads to a change of the van der Waals gap of only 3\% \cite{Hor:2010b}. {Currently, little is known about intercalation-induced changes of the near-surface structure of Bi$_2$Se$_3$ ,but such changes could be determined by surface x-ray diffraction, at least in principle.}

{The band bending picture for the explanation of the M-shaped states in Bi$_2$Se$_3$ has been questioned by recent results for the similar compound Bi$_2ti$Te$_3$ \cite{Chen:2012}. In Bi$_2$Te$_3$, similar M-shaped and distinct states in the valence band region are found upon exposing the sample to air or $N_2$, but the authors of Ref.  \cite{Chen:2012} argue that band bending cannot be the origin of these structures for two reasons: first, the structure is M-shaped and thus the effective mass changes from electron-like in the centre to hole-like on the wings. Thus, it is impossible for the entire state to be confined. Second, the band bending for confining the states would have to be larger than the total band width, which the authors determine to be around 300~meV. We believe these arguments are invalid. Concerning the first argument, the shape of the state in $k_{\parallel}$ is not relevant for the confinement that takes place in the  $z$ direction perpendicular to the surface. Along $k_z$, the most advanced GW-calculations show no large dispersion of the band, and it is not clear if it is hole-like or electron-like \cite{Yazyev:2012}. Concerning the second argument, the relevant bandwidth is also that in $k_z$, not in $k_{\parallel}$. Again, GW-theory predicts the width of the upper valence band to be very small  \cite{Yazyev:2012}, and even the band structure calculation shown in the supplementary material of Ref. \cite{Chen:2012} appears to give a large band width only because two bands are involved. Finally, it is not strictly necessary to have a band bending that exceeds the total band width in order to confine the valence band states. In a situation with a projected band gap below the valence band, the states with the lowest energy can already be confined by a much smaller band bending. A band bending larger than the total width is only required if the entire band is to be quantised.}




\section{Conclusion}

In conclusion, we have presented ARPES results for clean and adsorbate-covered  \bise. Different bulk dopings were studied. For carbon monoxide or rest gas adsorption, a strong downward bending of the near-surface bands was observed. This was found to lead to new spectral features in the energy region of both the valence band and the conduction band. These changes were explained by a confinement of the states and the formation of quantum well states. This interpretation is supported by simple calculations of the band bending effects and by photon energy-dependent ARPES measurements that show how the band dispersion in the direction perpendicular to the surface is lost and non-dispersing, sharp states appear within the energy regions of the projected bulk bands. For strong band bending, the conduction band quantum well states are strongly Rashba split. 
 
 \section{Acknowledgements}

We gratefully acknowledge financial support by the Lundbeck foundation, The Danish Council for Independent Research, Natural Sciences  and the Danish National Research Foundation as well as valuable discussions with P.~D.~C.~King, S.~V.~Eremeev and E.~V.~Chulkov.


\begin{thebibliography}{88}
\expandafter\ifx\csname natexlab\endcsname\relax\def\natexlab#1{#1}\fi
\expandafter\ifx\csname bibnamefont\endcsname\relax
  \def\bibnamefont#1{#1}\fi
\expandafter\ifx\csname bibfnamefont\endcsname\relax
  \def\bibfnamefont#1{#1}\fi
\expandafter\ifx\csname citenamefont\endcsname\relax
  \def\citenamefont#1{#1}\fi
\expandafter\ifx\csname url\endcsname\relax
  \def\url#1{\texttt{#1}}\fi
\expandafter\ifx\csname urlprefix\endcsname\relax\def\urlprefix{URL }\fi
\providecommand{\bibinfo}[2]{#2}
\providecommand{\eprint}[2][]{\url{#2}}

\bibitem[{\citenamefont{Klitzing et~al.}(1980)\citenamefont{Klitzing, Dorda,
  and Pepper}}]{Klitzing:1980}
\bibinfo{author}{\bibfnamefont{K.~v.} \bibnamefont{Klitzing}},
  \bibinfo{author}{\bibfnamefont{G.}~\bibnamefont{Dorda}}, \bibnamefont{and}
  \bibinfo{author}{\bibfnamefont{M.}~\bibnamefont{Pepper}},
  \bibinfo{journal}{Phys. Rev. Lett.} \textbf{\bibinfo{volume}{45}},
  \bibinfo{pages}{494} (\bibinfo{year}{1980}).

\bibitem[{\citenamefont{Kane and Mele}(2005{\natexlab{a}})}]{Kane:2005b}
\bibinfo{author}{\bibfnamefont{C.~L.} \bibnamefont{Kane}} \bibnamefont{and}
  \bibinfo{author}{\bibfnamefont{E.~J.} \bibnamefont{Mele}},
  \bibinfo{journal}{Physical Review Letters} \textbf{\bibinfo{volume}{95}},
  \bibinfo{eid}{146802} (pages~\bibinfo{numpages}{4})
  (\bibinfo{year}{2005}{\natexlab{a}}),
  \urlprefix\url{http://link.aps.org/abstract/PRL/v95/e146802}.

\bibitem[{\citenamefont{Kane and Mele}(2005{\natexlab{b}})}]{Kane:2005c}
\bibinfo{author}{\bibfnamefont{C.~L.} \bibnamefont{Kane}} \bibnamefont{and}
  \bibinfo{author}{\bibfnamefont{E.~J.} \bibnamefont{Mele}},
  \bibinfo{journal}{Physical Review Letters} \textbf{\bibinfo{volume}{95}},
  \bibinfo{eid}{226801} (pages~\bibinfo{numpages}{4})
  (\bibinfo{year}{2005}{\natexlab{b}}),
  \urlprefix\url{http://link.aps.org/abstract/PRL/v95/e226801}.

\bibitem[{\citenamefont{Bernevig and Zhang}(2006)}]{Bernevig:2006b}
\bibinfo{author}{\bibfnamefont{B.~A.} \bibnamefont{Bernevig}} \bibnamefont{and}
  \bibinfo{author}{\bibfnamefont{S.-C.} \bibnamefont{Zhang}},
  \bibinfo{journal}{Physical Review Letters} \textbf{\bibinfo{volume}{96}},
  \bibinfo{eid}{106802} (pages~\bibinfo{numpages}{4}) (\bibinfo{year}{2006}),
  \urlprefix\url{http://link.aps.org/abstract/PRL/v96/e106802}.

\bibitem[{\citenamefont{Bernevig et~al.}(2006)\citenamefont{Bernevig, Hughes,
  and Zhang}}]{Bernevig:2006}
\bibinfo{author}{\bibfnamefont{B.~A.} \bibnamefont{Bernevig}},
  \bibinfo{author}{\bibfnamefont{T.~L.} \bibnamefont{Hughes}},
  \bibnamefont{and} \bibinfo{author}{\bibfnamefont{S.-C.} \bibnamefont{Zhang}},
  \bibinfo{journal}{Science} \textbf{\bibinfo{volume}{314}},
  \bibinfo{pages}{1757} (\bibinfo{year}{2006}),
  \urlprefix\url{http://www.sciencemag.org/cgi/content/abstract/314/5806/1757}.

\bibitem[{\citenamefont{Moore and Balents}(2007)}]{Moore:2007}
\bibinfo{author}{\bibfnamefont{J.~E.} \bibnamefont{Moore}} \bibnamefont{and}
  \bibinfo{author}{\bibfnamefont{L.}~\bibnamefont{Balents}},
  \bibinfo{journal}{Phys. Rev. B} \textbf{\bibinfo{volume}{75}},
  \bibinfo{pages}{121306} (\bibinfo{year}{2007}).

\bibitem[{\citenamefont{Konig et~al.}(2007)\citenamefont{Konig, Wiedmann,
  Brune, Roth, Buhmann, Molenkamp, Qi, and Zhang}}]{Konig:2007}
\bibinfo{author}{\bibfnamefont{M.}~\bibnamefont{Konig}},
  \bibinfo{author}{\bibfnamefont{S.}~\bibnamefont{Wiedmann}},
  \bibinfo{author}{\bibfnamefont{C.}~\bibnamefont{Brune}},
  \bibinfo{author}{\bibfnamefont{A.}~\bibnamefont{Roth}},
  \bibinfo{author}{\bibfnamefont{H.}~\bibnamefont{Buhmann}},
  \bibinfo{author}{\bibfnamefont{L.~W.} \bibnamefont{Molenkamp}},
  \bibinfo{author}{\bibfnamefont{X.-L.} \bibnamefont{Qi}}, \bibnamefont{and}
  \bibinfo{author}{\bibfnamefont{S.-C.} \bibnamefont{Zhang}},
  \bibinfo{journal}{Science} \textbf{\bibinfo{volume}{318}},
  \bibinfo{pages}{766} (\bibinfo{year}{2007}),
  \urlprefix\url{http://www.sciencemag.org/cgi/content/abstract/318/5851/766}.

\bibitem[{\citenamefont{Zhang}(2008)}]{Zhang:2008}
\bibinfo{author}{\bibfnamefont{S.~C.} \bibnamefont{Zhang}},
  \bibinfo{journal}{Physics} \textbf{\bibinfo{volume}{1}}, \bibinfo{eid}{6}
  (\bibinfo{year}{2008}).

\bibitem[{\citenamefont{Moore}(2010)}]{Moore:2010}
\bibinfo{author}{\bibfnamefont{J.~E.} \bibnamefont{Moore}},
  \bibinfo{journal}{Nature} \textbf{\bibinfo{volume}{464}},
  \bibinfo{pages}{194} (\bibinfo{year}{2010}),
  \urlprefix\url{http://dx.doi.org/10.1038/nature08916}.

\bibitem[{\citenamefont{Hasan and Kane}(2010)}]{Hasan:2010}
\bibinfo{author}{\bibfnamefont{M.~Z.} \bibnamefont{Hasan}} \bibnamefont{and}
  \bibinfo{author}{\bibfnamefont{C.~L.} \bibnamefont{Kane}},
  \bibinfo{journal}{Rev. Mod. Phys.} \textbf{\bibinfo{volume}{82}},
  \bibinfo{pages}{3045} (\bibinfo{year}{2010}).

\bibitem[{\citenamefont{Barke et~al.}(2007)\citenamefont{Barke, Bennewitz,
  Crain, Erwin, Kirakosian, McChesney, and Himpsel}}]{Barke:2007a}
\bibinfo{author}{\bibfnamefont{I.}~\bibnamefont{Barke}},
  \bibinfo{author}{\bibfnamefont{R.}~\bibnamefont{Bennewitz}},
  \bibinfo{author}{\bibfnamefont{J.~N.} \bibnamefont{Crain}},
  \bibinfo{author}{\bibfnamefont{S.~C.} \bibnamefont{Erwin}},
  \bibinfo{author}{\bibfnamefont{A.}~\bibnamefont{Kirakosian}},
  \bibinfo{author}{\bibfnamefont{J.~L.} \bibnamefont{McChesney}},
  \bibnamefont{and} \bibinfo{author}{\bibfnamefont{F.~J.}
  \bibnamefont{Himpsel}}, \bibinfo{journal}{Solid State Communications}
  \textbf{\bibinfo{volume}{142}}, \bibinfo{pages}{617} (\bibinfo{year}{2007}),
  \urlprefix\url{http://www.sciencedirect.com/science/article/pii/S0038109807002761}.

\bibitem[{\citenamefont{Fu and Kane}(2007)}]{Fu:2007b}
\bibinfo{author}{\bibfnamefont{L.}~\bibnamefont{Fu}} \bibnamefont{and}
  \bibinfo{author}{\bibfnamefont{C.~L.} \bibnamefont{Kane}},
  \bibinfo{journal}{Physical Review B} \textbf{\bibinfo{volume}{76}},
  \bibinfo{eid}{045302} (pages~\bibinfo{numpages}{17}) (\bibinfo{year}{2007}),
  \urlprefix\url{http://link.aps.org/abstract/PRB/v76/e045302}.

\bibitem[{\citenamefont{Murakami}(2007)}]{Murakami:2007}
\bibinfo{author}{\bibfnamefont{S.}~\bibnamefont{Murakami}},
  \bibinfo{journal}{New Journal of Physics} \textbf{\bibinfo{volume}{9}},
  \bibinfo{pages}{356} (\bibinfo{year}{2007}),
  \urlprefix\url{http://stacks.iop.org/1367-2630/9/i=9/a=356}.

\bibitem[{\citenamefont{Hsieh et~al.}(2008)\citenamefont{Hsieh, Qian, Wray,
  Xia, Hor, Cava, and Hasan}}]{Hsieh:2008}
\bibinfo{author}{\bibfnamefont{D.}~\bibnamefont{Hsieh}},
  \bibinfo{author}{\bibfnamefont{D.}~\bibnamefont{Qian}},
  \bibinfo{author}{\bibfnamefont{L.}~\bibnamefont{Wray}},
  \bibinfo{author}{\bibfnamefont{Y.}~\bibnamefont{Xia}},
  \bibinfo{author}{\bibfnamefont{Y.~S.} \bibnamefont{Hor}},
  \bibinfo{author}{\bibfnamefont{R.~J.} \bibnamefont{Cava}}, \bibnamefont{and}
  \bibinfo{author}{\bibfnamefont{M.~Z.} \bibnamefont{Hasan}},
  \bibinfo{journal}{Nature} \textbf{\bibinfo{volume}{452}},
  \bibinfo{pages}{970} (\bibinfo{year}{2008}),
  \urlprefix\url{http://dx.doi.org/10.1038/nature06843}.

\bibitem[{\citenamefont{Ast and H{{\"o}}chst}(2001)}]{Ast:2001}
\bibinfo{author}{\bibfnamefont{C.~R.} \bibnamefont{Ast}} \bibnamefont{and}
  \bibinfo{author}{\bibfnamefont{H.}~\bibnamefont{H{{\"o}}chst}},
  \bibinfo{journal}{Physical Review Letters} \textbf{\bibinfo{volume}{87}},
  \bibinfo{pages}{177602} (\bibinfo{year}{2001}).

\bibitem[{\citenamefont{Zhang et~al.}(2009)\citenamefont{Zhang, Liu, Qi, Dai,
  Fang, and Zhang}}]{Zhang:2009}
\bibinfo{author}{\bibfnamefont{H.}~\bibnamefont{Zhang}},
  \bibinfo{author}{\bibfnamefont{C.-X.} \bibnamefont{Liu}},
  \bibinfo{author}{\bibfnamefont{X.-L.} \bibnamefont{Qi}},
  \bibinfo{author}{\bibfnamefont{X.}~\bibnamefont{Dai}},
  \bibinfo{author}{\bibfnamefont{Z.}~\bibnamefont{Fang}}, \bibnamefont{and}
  \bibinfo{author}{\bibfnamefont{S.-C.} \bibnamefont{Zhang}},
  \bibinfo{journal}{Nature Physics} \textbf{\bibinfo{volume}{5}},
  \bibinfo{pages}{438} (\bibinfo{year}{2009}),
  \urlprefix\url{http://dx.doi.org/10.1038/nphys1270}.

\bibitem[{\citenamefont{Xia et~al.}(2009)\citenamefont{Xia, Qian, Hsieh, Wray,
  Pal, Lin, Bansil, Grauer, Hor, Cava et~al.}}]{Xia:2009}
\bibinfo{author}{\bibfnamefont{Y.}~\bibnamefont{Xia}},
  \bibinfo{author}{\bibfnamefont{D.}~\bibnamefont{Qian}},
  \bibinfo{author}{\bibfnamefont{D.}~\bibnamefont{Hsieh}},
  \bibinfo{author}{\bibfnamefont{L.}~\bibnamefont{Wray}},
  \bibinfo{author}{\bibfnamefont{A.}~\bibnamefont{Pal}},
  \bibinfo{author}{\bibfnamefont{H.}~\bibnamefont{Lin}},
  \bibinfo{author}{\bibfnamefont{A.}~\bibnamefont{Bansil}},
  \bibinfo{author}{\bibfnamefont{D.}~\bibnamefont{Grauer}},
  \bibinfo{author}{\bibfnamefont{Y.~S.} \bibnamefont{Hor}},
  \bibinfo{author}{\bibfnamefont{R.~J.} \bibnamefont{Cava}},
  \bibnamefont{et~al.}, \bibinfo{journal}{Nature Physics}
  \textbf{\bibinfo{volume}{5}}, \bibinfo{pages}{398} (\bibinfo{year}{2009}),
  \urlprefix\url{http://dx.doi.org/10.1038/nphys1274}.

\bibitem[{\citenamefont{Noh et~al.}(2008)\citenamefont{Noh, Koh, Oh, Park, Kim,
  Rameau, Valla, Kidd, Johnson, Hu et~al.}}]{Noh:2008}
\bibinfo{author}{\bibfnamefont{H.-J.} \bibnamefont{Noh}},
  \bibinfo{author}{\bibfnamefont{H.}~\bibnamefont{Koh}},
  \bibinfo{author}{\bibfnamefont{S.-J.} \bibnamefont{Oh}},
  \bibinfo{author}{\bibfnamefont{J.-H.} \bibnamefont{Park}},
  \bibinfo{author}{\bibfnamefont{H.-D.} \bibnamefont{Kim}},
  \bibinfo{author}{\bibfnamefont{J.~D.} \bibnamefont{Rameau}},
  \bibinfo{author}{\bibfnamefont{T.}~\bibnamefont{Valla}},
  \bibinfo{author}{\bibfnamefont{T.~E.} \bibnamefont{Kidd}},
  \bibinfo{author}{\bibfnamefont{P.~D.} \bibnamefont{Johnson}},
  \bibinfo{author}{\bibfnamefont{Y.}~\bibnamefont{Hu}}, \bibnamefont{et~al.},
  \bibinfo{journal}{Europhysics Letters} \textbf{\bibinfo{volume}{81}},
  \bibinfo{pages}{57006} (\bibinfo{year}{2008}),
  \urlprefix\url{http://stacks.iop.org/0295-5075/81/i=5/a=57006}.

\bibitem[{\citenamefont{Chen et~al.}(2009)\citenamefont{Chen, Analytis, Chu,
  Liu, Mo, Qi, Zhang, Lu, Dai, Fang et~al.}}]{Chen:2009}
\bibinfo{author}{\bibfnamefont{Y.~L.} \bibnamefont{Chen}},
  \bibinfo{author}{\bibfnamefont{J.~G.} \bibnamefont{Analytis}},
  \bibinfo{author}{\bibfnamefont{J.~H.} \bibnamefont{Chu}},
  \bibinfo{author}{\bibfnamefont{Z.~K.} \bibnamefont{Liu}},
  \bibinfo{author}{\bibfnamefont{S.~K.} \bibnamefont{Mo}},
  \bibinfo{author}{\bibfnamefont{X.~L.} \bibnamefont{Qi}},
  \bibinfo{author}{\bibfnamefont{H.~J.} \bibnamefont{Zhang}},
  \bibinfo{author}{\bibfnamefont{D.~H.} \bibnamefont{Lu}},
  \bibinfo{author}{\bibfnamefont{X.}~\bibnamefont{Dai}},
  \bibinfo{author}{\bibfnamefont{Z.}~\bibnamefont{Fang}}, \bibnamefont{et~al.},
  \bibinfo{journal}{Science} \textbf{\bibinfo{volume}{325}},
  \bibinfo{pages}{178} (\bibinfo{year}{2009}),
  \urlprefix\url{http://www.sciencemag.org/cgi/content/abstract/325/5937/178}.

\bibitem[{\citenamefont{Hsieh et~al.}(2009{\natexlab{a}})\citenamefont{Hsieh,
  Xia, Qian, Wray, Dil, Meier, Osterwalder, Patthey, Checkelsky, Ong
  et~al.}}]{Hsieh:2009c}
\bibinfo{author}{\bibfnamefont{D.}~\bibnamefont{Hsieh}},
  \bibinfo{author}{\bibfnamefont{Y.}~\bibnamefont{Xia}},
  \bibinfo{author}{\bibfnamefont{D.}~\bibnamefont{Qian}},
  \bibinfo{author}{\bibfnamefont{L.}~\bibnamefont{Wray}},
  \bibinfo{author}{\bibfnamefont{J.~H.} \bibnamefont{Dil}},
  \bibinfo{author}{\bibfnamefont{F.}~\bibnamefont{Meier}},
  \bibinfo{author}{\bibfnamefont{J.}~\bibnamefont{Osterwalder}},
  \bibinfo{author}{\bibfnamefont{L.}~\bibnamefont{Patthey}},
  \bibinfo{author}{\bibfnamefont{J.~G.} \bibnamefont{Checkelsky}},
  \bibinfo{author}{\bibfnamefont{N.~P.} \bibnamefont{Ong}},
  \bibnamefont{et~al.}, \bibinfo{journal}{Nature}
  \textbf{\bibinfo{volume}{460}}, \bibinfo{pages}{1101}
  (\bibinfo{year}{2009}{\natexlab{a}}),
  \urlprefix\url{http://dx.doi.org/10.1038/nature08234}.

\bibitem[{\citenamefont{Hsieh et~al.}(2009{\natexlab{b}})\citenamefont{Hsieh,
  Xia, Wray, Qian, Pal, Dil, Osterwalder, Meier, Bihlmayer, Kane
  et~al.}}]{Hsieh:2009}
\bibinfo{author}{\bibfnamefont{D.}~\bibnamefont{Hsieh}},
  \bibinfo{author}{\bibfnamefont{Y.}~\bibnamefont{Xia}},
  \bibinfo{author}{\bibfnamefont{L.}~\bibnamefont{Wray}},
  \bibinfo{author}{\bibfnamefont{D.}~\bibnamefont{Qian}},
  \bibinfo{author}{\bibfnamefont{A.}~\bibnamefont{Pal}},
  \bibinfo{author}{\bibfnamefont{J.~H.} \bibnamefont{Dil}},
  \bibinfo{author}{\bibfnamefont{J.}~\bibnamefont{Osterwalder}},
  \bibinfo{author}{\bibfnamefont{F.}~\bibnamefont{Meier}},
  \bibinfo{author}{\bibfnamefont{G.}~\bibnamefont{Bihlmayer}},
  \bibinfo{author}{\bibfnamefont{C.~L.} \bibnamefont{Kane}},
  \bibnamefont{et~al.}, \bibinfo{journal}{Science}
  \textbf{\bibinfo{volume}{323}}, \bibinfo{pages}{919}
  (\bibinfo{year}{2009}{\natexlab{b}}),
  \urlprefix\url{http://www.sciencemag.org/cgi/content/abstract/323/5916/919}.

\bibitem[{\citenamefont{Hsieh et~al.}(2009{\natexlab{c}})\citenamefont{Hsieh,
  Xia, Qian, Wray, Meier, Dil, Osterwalder, Patthey, Fedorov, Lin
  et~al.}}]{Hsieh:2009b}
\bibinfo{author}{\bibfnamefont{D.}~\bibnamefont{Hsieh}},
  \bibinfo{author}{\bibfnamefont{Y.}~\bibnamefont{Xia}},
  \bibinfo{author}{\bibfnamefont{D.}~\bibnamefont{Qian}},
  \bibinfo{author}{\bibfnamefont{L.}~\bibnamefont{Wray}},
  \bibinfo{author}{\bibfnamefont{F.}~\bibnamefont{Meier}},
  \bibinfo{author}{\bibfnamefont{J.~H.} \bibnamefont{Dil}},
  \bibinfo{author}{\bibfnamefont{J.}~\bibnamefont{Osterwalder}},
  \bibinfo{author}{\bibfnamefont{L.}~\bibnamefont{Patthey}},
  \bibinfo{author}{\bibfnamefont{A.~V.} \bibnamefont{Fedorov}},
  \bibinfo{author}{\bibfnamefont{H.}~\bibnamefont{Lin}}, \bibnamefont{et~al.},
  \bibinfo{journal}{Physical Review Letters} \textbf{\bibinfo{volume}{103}},
  \bibinfo{eid}{146401} (pages~\bibinfo{numpages}{4})
  (\bibinfo{year}{2009}{\natexlab{c}}),
  \urlprefix\url{http://link.aps.org/abstract/PRL/v103/e146401}.

\bibitem[{\citenamefont{Roushan et~al.}(2009)\citenamefont{Roushan, Seo,
  Parker, Hor, Hsieh, Qian, Richardella, Hasan, Cava, and
  Yazdani}}]{Roushan:2009}
\bibinfo{author}{\bibfnamefont{P.}~\bibnamefont{Roushan}},
  \bibinfo{author}{\bibfnamefont{J.}~\bibnamefont{Seo}},
  \bibinfo{author}{\bibfnamefont{C.~V.} \bibnamefont{Parker}},
  \bibinfo{author}{\bibfnamefont{Y.~S.} \bibnamefont{Hor}},
  \bibinfo{author}{\bibfnamefont{D.}~\bibnamefont{Hsieh}},
  \bibinfo{author}{\bibfnamefont{D.}~\bibnamefont{Qian}},
  \bibinfo{author}{\bibfnamefont{A.}~\bibnamefont{Richardella}},
  \bibinfo{author}{\bibfnamefont{M.~Z.} \bibnamefont{Hasan}},
  \bibinfo{author}{\bibfnamefont{R.~J.} \bibnamefont{Cava}}, \bibnamefont{and}
  \bibinfo{author}{\bibfnamefont{A.}~\bibnamefont{Yazdani}},
  \bibinfo{journal}{Nature} \textbf{\bibinfo{volume}{460}},
  \bibinfo{pages}{1106} (\bibinfo{year}{2009}),
  \urlprefix\url{http://dx.doi.org/10.1038/nature08308}.

\bibitem[{\citenamefont{Alpichshev et~al.}(2010)\citenamefont{Alpichshev,
  Analytis, Chu, Fisher, Chen, Shen, Fang, and Kapitulnik}}]{Alpichshev:2010}
\bibinfo{author}{\bibfnamefont{Z.}~\bibnamefont{Alpichshev}},
  \bibinfo{author}{\bibfnamefont{J.~G.} \bibnamefont{Analytis}},
  \bibinfo{author}{\bibfnamefont{J.-H.} \bibnamefont{Chu}},
  \bibinfo{author}{\bibfnamefont{I.~R.} \bibnamefont{Fisher}},
  \bibinfo{author}{\bibfnamefont{Y.~L.} \bibnamefont{Chen}},
  \bibinfo{author}{\bibfnamefont{Z.~X.} \bibnamefont{Shen}},
  \bibinfo{author}{\bibfnamefont{A.}~\bibnamefont{Fang}}, \bibnamefont{and}
  \bibinfo{author}{\bibfnamefont{A.}~\bibnamefont{Kapitulnik}},
  \bibinfo{journal}{Phys. Rev. Lett.} \textbf{\bibinfo{volume}{104}},
  \bibinfo{pages}{016401} (\bibinfo{year}{2010}).

\bibitem[{\citenamefont{Petersen and Hedegard}(2000)}]{Petersen:2000}
\bibinfo{author}{\bibfnamefont{L.}~\bibnamefont{Petersen}} \bibnamefont{and}
  \bibinfo{author}{\bibfnamefont{P.}~\bibnamefont{Hedegard}},
  \bibinfo{journal}{Surface Science} \textbf{\bibinfo{volume}{459}},
  \bibinfo{pages}{49} (\bibinfo{year}{2000}).

\bibitem[{\citenamefont{Nechaev et~al.}(2009)\citenamefont{Nechaev, Jensen,
  Rienks, Silkin, Echenique, Chulkov, and Hofmann}}]{Nechaev:2009}
\bibinfo{author}{\bibfnamefont{I.~A.} \bibnamefont{Nechaev}},
  \bibinfo{author}{\bibfnamefont{M.~F.} \bibnamefont{Jensen}},
  \bibinfo{author}{\bibfnamefont{E.~D.~L.} \bibnamefont{Rienks}},
  \bibinfo{author}{\bibfnamefont{V.~M.} \bibnamefont{Silkin}},
  \bibinfo{author}{\bibfnamefont{P.~M.} \bibnamefont{Echenique}},
  \bibinfo{author}{\bibfnamefont{E.~V.} \bibnamefont{Chulkov}},
  \bibnamefont{and} \bibinfo{author}{\bibfnamefont{P.}~\bibnamefont{Hofmann}},
  \bibinfo{journal}{Physical Review B} \textbf{\bibinfo{volume}{80}},
  \bibinfo{eid}{113402} (\bibinfo{year}{2009}),
  \urlprefix\url{http://link.aps.org/abstract/PRB/v80/e113402}.

\bibitem[{\citenamefont{Gayone et~al.}(2003)\citenamefont{Gayone, Hoffmann, Li,
  and Hofmann}}]{Gayone:2003}
\bibinfo{author}{\bibfnamefont{J.~E.} \bibnamefont{Gayone}},
  \bibinfo{author}{\bibfnamefont{S.~V.} \bibnamefont{Hoffmann}},
  \bibinfo{author}{\bibfnamefont{Z.}~\bibnamefont{Li}}, \bibnamefont{and}
  \bibinfo{author}{\bibfnamefont{P.}~\bibnamefont{Hofmann}},
  \bibinfo{journal}{Physical Review Letters} \textbf{\bibinfo{volume}{91}},
  \bibinfo{pages}{127601} (\bibinfo{year}{2003}).

\bibitem[{\citenamefont{Hofmann}(2006)}]{Hofmann:2006}
\bibinfo{author}{\bibfnamefont{P.}~\bibnamefont{Hofmann}},
  \bibinfo{journal}{Progress in Surface Science} \textbf{\bibinfo{volume}{81}},
  \bibinfo{pages}{191} (\bibinfo{year}{2006}), ISSN \bibinfo{issn}{5}.

\bibitem[{\citenamefont{Kim et~al.}(2005)\citenamefont{Kim, Wells, Kirkegaard,
  Li, Hoffmann, Gayone, Fernandez-Torrente, Haberle, Pascual, Moore
  et~al.}}]{Kim:2005b}
\bibinfo{author}{\bibfnamefont{T.~K.} \bibnamefont{Kim}},
  \bibinfo{author}{\bibfnamefont{J.}~\bibnamefont{Wells}},
  \bibinfo{author}{\bibfnamefont{C.}~\bibnamefont{Kirkegaard}},
  \bibinfo{author}{\bibfnamefont{Z.}~\bibnamefont{Li}},
  \bibinfo{author}{\bibfnamefont{S.~V.} \bibnamefont{Hoffmann}},
  \bibinfo{author}{\bibfnamefont{J.~E.} \bibnamefont{Gayone}},
  \bibinfo{author}{\bibfnamefont{I.}~\bibnamefont{Fernandez-Torrente}},
  \bibinfo{author}{\bibfnamefont{P.}~\bibnamefont{Haberle}},
  \bibinfo{author}{\bibfnamefont{J.~I.} \bibnamefont{Pascual}},
  \bibinfo{author}{\bibfnamefont{K.~T.} \bibnamefont{Moore}},
  \bibnamefont{et~al.}, \bibinfo{journal}{Physical Review B}
  \textbf{\bibinfo{volume}{72}}, \bibinfo{eid}{085440}
  (pages~\bibinfo{numpages}{7}) (\bibinfo{year}{2005}),
  \urlprefix\url{http://link.aps.org/abstract/PRB/v72/e085440}.

\bibitem[{\citenamefont{Koroteev et~al.}(2004)\citenamefont{Koroteev,
  Bihlmayer, Gayone, Chulkov, Bl{\"u}gel, Echenique, and
  Hofmann}}]{Koroteev:2004}
\bibinfo{author}{\bibfnamefont{Y.~M.} \bibnamefont{Koroteev}},
  \bibinfo{author}{\bibfnamefont{G.}~\bibnamefont{Bihlmayer}},
  \bibinfo{author}{\bibfnamefont{J.~E.} \bibnamefont{Gayone}},
  \bibinfo{author}{\bibfnamefont{E.~V.} \bibnamefont{Chulkov}},
  \bibinfo{author}{\bibfnamefont{S.}~\bibnamefont{Bl{\"u}gel}},
  \bibinfo{author}{\bibfnamefont{P.~M.} \bibnamefont{Echenique}},
  \bibnamefont{and} \bibinfo{author}{\bibfnamefont{P.}~\bibnamefont{Hofmann}},
  \bibinfo{journal}{Physical Review Letters} \textbf{\bibinfo{volume}{93}},
  \bibinfo{pages}{046403} (\bibinfo{year}{2004}).

\bibitem[{\citenamefont{Pascual et~al.}(2004)\citenamefont{Pascual, Bihlmayer,
  Koroteev, Rust, Ceballos, Hansmann, Horn, Chulkov, Blugel, Echenique
  et~al.}}]{Pascual:2004}
\bibinfo{author}{\bibfnamefont{J.~I.} \bibnamefont{Pascual}},
  \bibinfo{author}{\bibfnamefont{G.}~\bibnamefont{Bihlmayer}},
  \bibinfo{author}{\bibfnamefont{Y.~M.} \bibnamefont{Koroteev}},
  \bibinfo{author}{\bibfnamefont{H.~P.} \bibnamefont{Rust}},
  \bibinfo{author}{\bibfnamefont{G.}~\bibnamefont{Ceballos}},
  \bibinfo{author}{\bibfnamefont{M.}~\bibnamefont{Hansmann}},
  \bibinfo{author}{\bibfnamefont{K.}~\bibnamefont{Horn}},
  \bibinfo{author}{\bibfnamefont{E.~V.} \bibnamefont{Chulkov}},
  \bibinfo{author}{\bibfnamefont{S.}~\bibnamefont{Blugel}},
  \bibinfo{author}{\bibfnamefont{P.~M.} \bibnamefont{Echenique}},
  \bibnamefont{et~al.}, \bibinfo{journal}{Physical Review Letters}
  \textbf{\bibinfo{volume}{93}}, \bibinfo{pages}{196802}
  (\bibinfo{year}{2004}).

\bibitem[{\citenamefont{Teo et~al.}(2008)\citenamefont{Teo, Fu, and
  Kane}}]{Teo:2008}
\bibinfo{author}{\bibfnamefont{J.~C.~Y.} \bibnamefont{Teo}},
  \bibinfo{author}{\bibfnamefont{L.}~\bibnamefont{Fu}}, \bibnamefont{and}
  \bibinfo{author}{\bibfnamefont{C.~L.} \bibnamefont{Kane}},
  \bibinfo{journal}{Physical Review B} \textbf{\bibinfo{volume}{78}},
  \bibinfo{eid}{045426} (pages~\bibinfo{numpages}{15}) (\bibinfo{year}{2008}),
  \urlprefix\url{http://link.aps.org/abstract/PRB/v78/e045426}.

\bibitem[{\citenamefont{Eremeev et~al.}(2010)\citenamefont{Eremeev, Koroteev,
  and Chulkov}}]{Eremeev:2010b}
\bibinfo{author}{\bibfnamefont{S.}~\bibnamefont{Eremeev}},
  \bibinfo{author}{\bibfnamefont{Y.}~\bibnamefont{Koroteev}}, \bibnamefont{and}
  \bibinfo{author}{\bibfnamefont{E.}~\bibnamefont{Chulkov}},
  \bibinfo{journal}{JETP Letters} \textbf{\bibinfo{volume}{91}},
  \bibinfo{pages}{387} (\bibinfo{year}{2010}), ISSN \bibinfo{issn}{0021-3640},
  \bibinfo{note}{10.1134/S0021364010080059},
  \urlprefix\url{http://dx.doi.org/10.1134/S0021364010080059}.

\bibitem[{\citenamefont{Bianchi
  et~al.}(2010{\natexlab{a}})\citenamefont{Bianchi, Guan, Bao, Mi, Iversen,
  King, and Hofmann}}]{Bianchi:2010b}
\bibinfo{author}{\bibfnamefont{M.}~\bibnamefont{Bianchi}},
  \bibinfo{author}{\bibfnamefont{D.}~\bibnamefont{Guan}},
  \bibinfo{author}{\bibfnamefont{S.}~\bibnamefont{Bao}},
  \bibinfo{author}{\bibfnamefont{J.}~\bibnamefont{Mi}},
  \bibinfo{author}{\bibfnamefont{B.~B.} \bibnamefont{Iversen}},
  \bibinfo{author}{\bibfnamefont{P.~D.~C.} \bibnamefont{King}},
  \bibnamefont{and} \bibinfo{author}{\bibfnamefont{P.}~\bibnamefont{Hofmann}},
  \bibinfo{journal}{Nature Communications} \textbf{\bibinfo{volume}{1}},
  \bibinfo{pages}{128} (\bibinfo{year}{2010}{\natexlab{a}}).

\bibitem[{\citenamefont{Bychkov and Rashba}(1984)}]{Bychkov:1984b}
\bibinfo{author}{\bibfnamefont{Y.~A.} \bibnamefont{Bychkov}} \bibnamefont{and}
  \bibinfo{author}{\bibfnamefont{E.~I.} \bibnamefont{Rashba}},
  \bibinfo{journal}{JETP Letters} \textbf{\bibinfo{volume}{{39}}},
  \bibinfo{pages}{78} (\bibinfo{year}{1984}).

\bibitem[{\citenamefont{LaShell et~al.}(1996)\citenamefont{LaShell, McDougall,
  and Jensen}}]{Lashell:1996}
\bibinfo{author}{\bibfnamefont{S.}~\bibnamefont{LaShell}},
  \bibinfo{author}{\bibfnamefont{B.~A.} \bibnamefont{McDougall}},
  \bibnamefont{and} \bibinfo{author}{\bibfnamefont{E.}~\bibnamefont{Jensen}},
  \bibinfo{journal}{Physical Review Letters} \textbf{\bibinfo{volume}{77}},
  \bibinfo{pages}{3419} (\bibinfo{year}{1996}).

\bibitem[{\citenamefont{Agergaard et~al.}(2001)\citenamefont{Agergaard,
  S{\o}ndergaard, Li, Nielsen, Hoffmann, Li, and Hofmann}}]{Agergaard:2001}
\bibinfo{author}{\bibfnamefont{S.}~\bibnamefont{Agergaard}},
  \bibinfo{author}{\bibfnamefont{C.}~\bibnamefont{S{\o}ndergaard}},
  \bibinfo{author}{\bibfnamefont{H.}~\bibnamefont{Li}},
  \bibinfo{author}{\bibfnamefont{M.~B.} \bibnamefont{Nielsen}},
  \bibinfo{author}{\bibfnamefont{S.~V.} \bibnamefont{Hoffmann}},
  \bibinfo{author}{\bibfnamefont{Z.}~\bibnamefont{Li}}, \bibnamefont{and}
  \bibinfo{author}{\bibfnamefont{P.}~\bibnamefont{Hofmann}},
  \bibinfo{journal}{New Journal of Physics} \textbf{\bibinfo{volume}{3}},
  \bibinfo{pages}{15.1} (\bibinfo{year}{2001}).

\bibitem[{\citenamefont{Ast et~al.}(2007)\citenamefont{Ast, Henk, Ernst,
  Moreschini, Falub, Pacile, Bruno, Kern, and Grioni}}]{Ast:2007}
\bibinfo{author}{\bibfnamefont{C.~R.} \bibnamefont{Ast}},
  \bibinfo{author}{\bibfnamefont{J.}~\bibnamefont{Henk}},
  \bibinfo{author}{\bibfnamefont{A.}~\bibnamefont{Ernst}},
  \bibinfo{author}{\bibfnamefont{L.}~\bibnamefont{Moreschini}},
  \bibinfo{author}{\bibfnamefont{M.~C.} \bibnamefont{Falub}},
  \bibinfo{author}{\bibfnamefont{D.}~\bibnamefont{Pacile}},
  \bibinfo{author}{\bibfnamefont{P.}~\bibnamefont{Bruno}},
  \bibinfo{author}{\bibfnamefont{K.}~\bibnamefont{Kern}}, \bibnamefont{and}
  \bibinfo{author}{\bibfnamefont{M.}~\bibnamefont{Grioni}},
  \bibinfo{journal}{Physical Review Letters} \textbf{\bibinfo{volume}{98}},
  \bibinfo{pages}{186807} (\bibinfo{year}{2007}).

\bibitem[{\citenamefont{Wray et~al.}(2011)\citenamefont{Wray, Xu, Xia, Hsieh,
  Fedorov, Hor, Cava, Bansil, Lin, and Hasan}}]{Wray:2011}
\bibinfo{author}{\bibfnamefont{L.~A.} \bibnamefont{Wray}},
  \bibinfo{author}{\bibfnamefont{S.-Y.} \bibnamefont{Xu}},
  \bibinfo{author}{\bibfnamefont{Y.}~\bibnamefont{Xia}},
  \bibinfo{author}{\bibfnamefont{D.}~\bibnamefont{Hsieh}},
  \bibinfo{author}{\bibfnamefont{A.~V.} \bibnamefont{Fedorov}},
  \bibinfo{author}{\bibfnamefont{Y.~S.} \bibnamefont{Hor}},
  \bibinfo{author}{\bibfnamefont{R.~J.} \bibnamefont{Cava}},
  \bibinfo{author}{\bibfnamefont{A.}~\bibnamefont{Bansil}},
  \bibinfo{author}{\bibfnamefont{H.}~\bibnamefont{Lin}}, \bibnamefont{and}
  \bibinfo{author}{\bibfnamefont{M.~Z.} \bibnamefont{Hasan}},
  \bibinfo{journal}{Nature Physics} \textbf{\bibinfo{volume}{7}},
  \bibinfo{pages}{32} (\bibinfo{year}{2011}),
  \urlprefix\url{http://dx.doi.org/10.1038/nphys1838}.

\bibitem[{\citenamefont{King et~al.}(2011)\citenamefont{King, Hatch, Bianchi,
  Ovsyannikov, Lupulescu, Landolt, Slomski, Dil, Guan, Mi et~al.}}]{King:2011}
\bibinfo{author}{\bibfnamefont{P.~D.~C.} \bibnamefont{King}},
  \bibinfo{author}{\bibfnamefont{R.~C.} \bibnamefont{Hatch}},
  \bibinfo{author}{\bibfnamefont{M.}~\bibnamefont{Bianchi}},
  \bibinfo{author}{\bibfnamefont{R.}~\bibnamefont{Ovsyannikov}},
  \bibinfo{author}{\bibfnamefont{C.}~\bibnamefont{Lupulescu}},
  \bibinfo{author}{\bibfnamefont{G.}~\bibnamefont{Landolt}},
  \bibinfo{author}{\bibfnamefont{B.}~\bibnamefont{Slomski}},
  \bibinfo{author}{\bibfnamefont{J.~H.} \bibnamefont{Dil}},
  \bibinfo{author}{\bibfnamefont{D.}~\bibnamefont{Guan}},
  \bibinfo{author}{\bibfnamefont{J.~L.} \bibnamefont{Mi}},
  \bibnamefont{et~al.}, \bibinfo{journal}{Phys. Rev. Lett.}
  \textbf{\bibinfo{volume}{107}}, \bibinfo{pages}{096802}
  (\bibinfo{year}{2011}).

\bibitem[{\citenamefont{Bianchi et~al.}(2011)\citenamefont{Bianchi, Hatch, Mi,
  Iversen, and Hofmann}}]{Bianchi:2011}
\bibinfo{author}{\bibfnamefont{M.}~\bibnamefont{Bianchi}},
  \bibinfo{author}{\bibfnamefont{R.~C.} \bibnamefont{Hatch}},
  \bibinfo{author}{\bibfnamefont{J.}~\bibnamefont{Mi}},
  \bibinfo{author}{\bibfnamefont{B.~B.} \bibnamefont{Iversen}},
  \bibnamefont{and} \bibinfo{author}{\bibfnamefont{P.}~\bibnamefont{Hofmann}},
  \bibinfo{journal}{Phys. Rev. Lett.} \textbf{\bibinfo{volume}{107}},
  \bibinfo{pages}{086802} (\bibinfo{year}{2011}).

\bibitem[{\citenamefont{Benia et~al.}(2011)\citenamefont{Benia, Lin, Kern, and
  Ast}}]{Benia:2011}
\bibinfo{author}{\bibfnamefont{H.~M.} \bibnamefont{Benia}},
  \bibinfo{author}{\bibfnamefont{C.}~\bibnamefont{Lin}},
  \bibinfo{author}{\bibfnamefont{K.}~\bibnamefont{Kern}}, \bibnamefont{and}
  \bibinfo{author}{\bibfnamefont{C.~R.} \bibnamefont{Ast}},
  \bibinfo{journal}{Phys. Rev. Lett.} \textbf{\bibinfo{volume}{107}},
  \bibinfo{pages}{177602} (\bibinfo{year}{2011}),
  \urlprefix\url{http://link.aps.org/doi/10.1103/PhysRevLett.107.177602}.

\bibitem[{\citenamefont{Pan et~al.}(2011)\citenamefont{Pan, Gardner, Chu, Lee,
  and Valla}}]{Pan:2011a}
\bibinfo{author}{\bibfnamefont{Z.-H.} \bibnamefont{Pan}},
  \bibinfo{author}{\bibfnamefont{D.~R.} \bibnamefont{Gardner}},
  \bibinfo{author}{\bibfnamefont{S.}~\bibnamefont{Chu}},
  \bibinfo{author}{\bibfnamefont{Y.~S.} \bibnamefont{Lee}}, \bibnamefont{and}
  \bibinfo{author}{\bibfnamefont{T.}~\bibnamefont{Valla}},
  \bibinfo{journal}{arXiv:1104.0966v1}  (\bibinfo{year}{2011}).

\bibitem[{\citenamefont{Valla et~al.}(2012)\citenamefont{Valla, Pan, Gardner,
  Lee, and Chu}}]{Valla:2012}
\bibinfo{author}{\bibfnamefont{T.}~\bibnamefont{Valla}},
  \bibinfo{author}{\bibfnamefont{Z.-H.} \bibnamefont{Pan}},
  \bibinfo{author}{\bibfnamefont{D.}~\bibnamefont{Gardner}},
  \bibinfo{author}{\bibfnamefont{Y.~S.} \bibnamefont{Lee}}, \bibnamefont{and}
  \bibinfo{author}{\bibfnamefont{S.}~\bibnamefont{Chu}},
  \bibinfo{journal}{Phys. Rev. Lett.} \textbf{\bibinfo{volume}{108}},
  \bibinfo{pages}{117601} (\bibinfo{year}{2012}),
  \urlprefix\url{http://link.aps.org/doi/10.1103/PhysRevLett.108.117601}.

\bibitem[{\citenamefont{Bianchi
  et~al.}(2010{\natexlab{b}})\citenamefont{Bianchi, Rienks, Lizzit, Baraldi,
  Balog, Hornek\ae{}r, and Hofmann}}]{Bianchi:2010}
\bibinfo{author}{\bibfnamefont{M.}~\bibnamefont{Bianchi}},
  \bibinfo{author}{\bibfnamefont{E.~D.~L.} \bibnamefont{Rienks}},
  \bibinfo{author}{\bibfnamefont{S.}~\bibnamefont{Lizzit}},
  \bibinfo{author}{\bibfnamefont{A.}~\bibnamefont{Baraldi}},
  \bibinfo{author}{\bibfnamefont{R.}~\bibnamefont{Balog}},
  \bibinfo{author}{\bibfnamefont{L.}~\bibnamefont{Hornek\ae{}r}},
  \bibnamefont{and} \bibinfo{author}{\bibfnamefont{P.}~\bibnamefont{Hofmann}},
  \bibinfo{journal}{Phys. Rev. B} \textbf{\bibinfo{volume}{81}},
  \bibinfo{pages}{041403} (\bibinfo{year}{2010}{\natexlab{b}}).

\bibitem[{\citenamefont{{Eremeev} et~al.}(2011)\citenamefont{{Eremeev},
  {Menshchikova}, {Vergniory}, and {Chulkov}}}]{Eremeev:2011b}
\bibinfo{author}{\bibfnamefont{S.~V.} \bibnamefont{{Eremeev}}},
  \bibinfo{author}{\bibfnamefont{T.~V.} \bibnamefont{{Menshchikova}}},
  \bibinfo{author}{\bibfnamefont{M.~G.} \bibnamefont{{Vergniory}}},
  \bibnamefont{and} \bibinfo{author}{\bibfnamefont{E.~V.}
  \bibnamefont{{Chulkov}}}, \bibinfo{journal}{ArXiv e-prints 1107.3208}
  (\bibinfo{year}{2011}), \eprint{1107.3208}.

\bibitem[{\citenamefont{Menshchikova et~al.}(2011)\citenamefont{Menshchikova,
  Eremeev, and Chulkov}}]{Menshchikova:2011}
\bibinfo{author}{\bibfnamefont{T.~V.} \bibnamefont{Menshchikova}},
  \bibinfo{author}{\bibfnamefont{S.~V.} \bibnamefont{Eremeev}},
  \bibnamefont{and} \bibinfo{author}{\bibfnamefont{E.~V.}
  \bibnamefont{Chulkov}}, \bibinfo{journal}{JETP Letters}
  \textbf{\bibinfo{volume}{94}}, \bibinfo{pages}{106} (\bibinfo{year}{2011}).

\bibitem[{\citenamefont{Ye et~al.}(2011)\citenamefont{Ye, Eremeev, Kuroda,
  Nakatake, Kim, Yamada, Krasovskii, Chulkov, Arita, Miyahara
  et~al.}}]{Ye:2011}
\bibinfo{author}{\bibfnamefont{M.}~\bibnamefont{Ye}},
  \bibinfo{author}{\bibfnamefont{S.~V.} \bibnamefont{Eremeev}},
  \bibinfo{author}{\bibfnamefont{K.}~\bibnamefont{Kuroda}},
  \bibinfo{author}{\bibfnamefont{M.}~\bibnamefont{Nakatake}},
  \bibinfo{author}{\bibfnamefont{S.}~\bibnamefont{Kim}},
  \bibinfo{author}{\bibfnamefont{Y.}~\bibnamefont{Yamada}},
  \bibinfo{author}{\bibfnamefont{E.~E.} \bibnamefont{Krasovskii}},
  \bibinfo{author}{\bibfnamefont{E.~V.} \bibnamefont{Chulkov}},
  \bibinfo{author}{\bibfnamefont{M.}~\bibnamefont{Arita}},
  \bibinfo{author}{\bibfnamefont{H.}~\bibnamefont{Miyahara}},
  \bibnamefont{et~al.}, \bibinfo{journal}{ArXiv e-prints 1112.5869}
  (\bibinfo{year}{2011}), \eprint{arXiv/1112.5869}.

\bibitem[{\citenamefont{Himpsel and Smith}(1985)}]{Himpsel:1985}
\bibinfo{author}{\bibfnamefont{F.~J.} \bibnamefont{Himpsel}} \bibnamefont{and}
  \bibinfo{author}{\bibfnamefont{N.~V.} \bibnamefont{Smith}},
  \bibinfo{journal}{Physics Today} pp. \bibinfo{pages}{60--66}
  (\bibinfo{year}{1985}).

\bibitem[{\citenamefont{Plummer and Eberhardt}(1982)}]{Plummer:1982}
\bibinfo{author}{\bibfnamefont{E.~W.} \bibnamefont{Plummer}} \bibnamefont{and}
  \bibinfo{author}{\bibfnamefont{W.}~\bibnamefont{Eberhardt}},
  \bibinfo{journal}{Advances in Chemical Physics}
  \textbf{\bibinfo{volume}{49}}, \bibinfo{pages}{533} (\bibinfo{year}{1982}).

\bibitem[{\citenamefont{Kevan}(1992)}]{Kevan:1992}
\bibinfo{editor}{\bibfnamefont{S.~D.} \bibnamefont{Kevan}}, ed.,
  \emph{\bibinfo{title}{Angle-resolved photoemission}},
  vol.~\bibinfo{volume}{74} of \emph{\bibinfo{series}{Studies in Surface
  Chemistry and Catalysis}} (\bibinfo{publisher}{Elsevier},
  \bibinfo{address}{Amsterdam}, \bibinfo{year}{1992}).

\bibitem[{\citenamefont{Matzdorf}(1998)}]{Matzdorf:1998}
\bibinfo{author}{\bibfnamefont{R.}~\bibnamefont{Matzdorf}},
  \bibinfo{journal}{Surface Science Reports} \textbf{\bibinfo{volume}{30}},
  \bibinfo{pages}{153} (\bibinfo{year}{1998}).

\bibitem[{\citenamefont{H{{\"u}}fner}(2003)}]{Hufner:2003}
\bibinfo{author}{\bibfnamefont{S.}~\bibnamefont{H{{\"u}}fner}},
  \emph{\bibinfo{title}{Photoelectron spectroscopy}}
  (\bibinfo{publisher}{Springer}, \bibinfo{address}{Berlin},
  \bibinfo{year}{2003}), \bibinfo{edition}{3rd} ed.

\bibitem[{\citenamefont{Hofmann et~al.}(2009)\citenamefont{Hofmann, Sklyadneva,
  Rienks, and Chulkov}}]{Hofmann:2009b}
\bibinfo{author}{\bibfnamefont{P.}~\bibnamefont{Hofmann}},
  \bibinfo{author}{\bibfnamefont{I.~Y.} \bibnamefont{Sklyadneva}},
  \bibinfo{author}{\bibfnamefont{E.~D.~L.} \bibnamefont{Rienks}},
  \bibnamefont{and} \bibinfo{author}{\bibfnamefont{E.~V.}
  \bibnamefont{Chulkov}}, \bibinfo{journal}{New Journal of Physics}
  \textbf{\bibinfo{volume}{11}}, \bibinfo{pages}{125005}
  (\bibinfo{year}{2009}),
  \urlprefix\url{http://stacks.iop.org/1367-2630/11/i=12/a=125005}.

\bibitem[{\citenamefont{Tamm}(1932)}]{Tamm:1932}
\bibinfo{author}{\bibfnamefont{I.}~\bibnamefont{Tamm}}, \bibinfo{journal}{Phys.
  Z. Sowjetunion} \textbf{\bibinfo{volume}{1}} (\bibinfo{year}{1932}).

\bibitem[{\citenamefont{Shockley}(1939)}]{Shockley:1939}
\bibinfo{author}{\bibfnamefont{W.}~\bibnamefont{Shockley}},
  \bibinfo{journal}{Phys. Rev.} \textbf{\bibinfo{volume}{56}},
  \bibinfo{pages}{317} (\bibinfo{year}{1939}).

\bibitem[{\citenamefont{L{\"u}th}(1992)}]{Luth:1992}
\bibinfo{author}{\bibfnamefont{H.}~\bibnamefont{L{\"u}th}},
  \emph{\bibinfo{title}{Surfaces and interfaces of solid materials}}
  (\bibinfo{publisher}{Springer}, \bibinfo{year}{1992}),
  \bibinfo{edition}{third edition} ed.

\bibitem[{\citenamefont{Davison and
  St{\c{e}}{\'{s}}licka}(1992)}]{Davison:1992}
\bibinfo{author}{\bibfnamefont{S.~G.} \bibnamefont{Davison}} \bibnamefont{and}
  \bibinfo{author}{\bibfnamefont{M.}~\bibnamefont{St{\c{e}}{\'{s}}licka}},
  \emph{\bibinfo{title}{Basic theory of surface states}},
  vol.~\bibinfo{volume}{46} of \emph{\bibinfo{series}{Monographs on the Physics
  and Chemistry of Materials}} (\bibinfo{publisher}{Oxford University Press},
  \bibinfo{address}{Oxford}, \bibinfo{year}{1992}).

\bibitem[{\citenamefont{Colakerol et~al.}(2006)\citenamefont{Colakerol, Veal,
  Jeong, Plucinski, DeMasi, Learmonth, Glans, Wang, Zhang, Piper
  et~al.}}]{Colakerol:2006}
\bibinfo{author}{\bibfnamefont{L.}~\bibnamefont{Colakerol}},
  \bibinfo{author}{\bibfnamefont{T.~D.} \bibnamefont{Veal}},
  \bibinfo{author}{\bibfnamefont{H.-K.} \bibnamefont{Jeong}},
  \bibinfo{author}{\bibfnamefont{L.}~\bibnamefont{Plucinski}},
  \bibinfo{author}{\bibfnamefont{A.}~\bibnamefont{DeMasi}},
  \bibinfo{author}{\bibfnamefont{T.}~\bibnamefont{Learmonth}},
  \bibinfo{author}{\bibfnamefont{P.-A.} \bibnamefont{Glans}},
  \bibinfo{author}{\bibfnamefont{S.}~\bibnamefont{Wang}},
  \bibinfo{author}{\bibfnamefont{Y.}~\bibnamefont{Zhang}},
  \bibinfo{author}{\bibfnamefont{L.~F.~J.} \bibnamefont{Piper}},
  \bibnamefont{et~al.}, \bibinfo{journal}{Physical Review Letters}
  \textbf{\bibinfo{volume}{97}}, \bibinfo{eid}{237601}
  (pages~\bibinfo{numpages}{4}) (\bibinfo{year}{2006}),
  \urlprefix\url{http://link.aps.org/abstract/PRL/v97/e237601}.

\bibitem[{\citenamefont{Piper et~al.}(2008)\citenamefont{Piper, Colakerol,
  King, Schleife, Zuniga-Perez, Glans, Learmonth, Federov, Veal, Fuchs
  et~al.}}]{Piper:2008}
\bibinfo{author}{\bibfnamefont{L.~F.~J.} \bibnamefont{Piper}},
  \bibinfo{author}{\bibfnamefont{L.}~\bibnamefont{Colakerol}},
  \bibinfo{author}{\bibfnamefont{P.~D.~C.} \bibnamefont{King}},
  \bibinfo{author}{\bibfnamefont{A.}~\bibnamefont{Schleife}},
  \bibinfo{author}{\bibfnamefont{J.}~\bibnamefont{Zuniga-Perez}},
  \bibinfo{author}{\bibfnamefont{P.-A.} \bibnamefont{Glans}},
  \bibinfo{author}{\bibfnamefont{T.}~\bibnamefont{Learmonth}},
  \bibinfo{author}{\bibfnamefont{A.}~\bibnamefont{Federov}},
  \bibinfo{author}{\bibfnamefont{T.~D.} \bibnamefont{Veal}},
  \bibinfo{author}{\bibfnamefont{F.}~\bibnamefont{Fuchs}},
  \bibnamefont{et~al.}, \bibinfo{journal}{Physical Review B}
  \textbf{\bibinfo{volume}{78}}, \bibinfo{eid}{165127}
  (pages~\bibinfo{numpages}{5}) (\bibinfo{year}{2008}),
  \urlprefix\url{http://link.aps.org/abstract/PRB/v78/e165127}.

\bibitem[{\citenamefont{King et~al.}(2008{\natexlab{a}})\citenamefont{King,
  Veal, and McConville}}]{King:2008}
\bibinfo{author}{\bibfnamefont{P.~D.~C.} \bibnamefont{King}},
  \bibinfo{author}{\bibfnamefont{T.~D.} \bibnamefont{Veal}}, \bibnamefont{and}
  \bibinfo{author}{\bibfnamefont{C.~F.} \bibnamefont{McConville}},
  \bibinfo{journal}{Physical Review B} \textbf{\bibinfo{volume}{77}},
  \bibinfo{eid}{125305} (pages~\bibinfo{numpages}{7})
  (\bibinfo{year}{2008}{\natexlab{a}}),
  \urlprefix\url{http://link.aps.org/abstract/PRB/v77/e125305}.

\bibitem[{\citenamefont{King et~al.}(2008{\natexlab{b}})\citenamefont{King,
  Veal, Payne, Bourlange, Egdell, and McConville}}]{King:2008c}
\bibinfo{author}{\bibfnamefont{P.~D.~C.} \bibnamefont{King}},
  \bibinfo{author}{\bibfnamefont{T.~D.} \bibnamefont{Veal}},
  \bibinfo{author}{\bibfnamefont{D.~J.} \bibnamefont{Payne}},
  \bibinfo{author}{\bibfnamefont{A.}~\bibnamefont{Bourlange}},
  \bibinfo{author}{\bibfnamefont{R.~G.} \bibnamefont{Egdell}},
  \bibnamefont{and} \bibinfo{author}{\bibfnamefont{C.~F.}
  \bibnamefont{McConville}}, \bibinfo{journal}{Phys. Rev. Lett.}
  \textbf{\bibinfo{volume}{101}}, \bibinfo{pages}{116808}
  (\bibinfo{year}{2008}{\natexlab{b}}).

\bibitem[{\citenamefont{King et~al.}(2010)\citenamefont{King, Veal, McConville,
  Z\'u\~niga P\'erez, Mu\~noz Sanjos\'e, Hopkinson, Rienks, Jensen, and
  Hofmann}}]{King:2010}
\bibinfo{author}{\bibfnamefont{P.~D.~C.} \bibnamefont{King}},
  \bibinfo{author}{\bibfnamefont{T.~D.} \bibnamefont{Veal}},
  \bibinfo{author}{\bibfnamefont{C.~F.} \bibnamefont{McConville}},
  \bibinfo{author}{\bibfnamefont{J.}~\bibnamefont{Z\'u\~niga P\'erez}},
  \bibinfo{author}{\bibfnamefont{V.}~\bibnamefont{Mu\~noz Sanjos\'e}},
  \bibinfo{author}{\bibfnamefont{M.}~\bibnamefont{Hopkinson}},
  \bibinfo{author}{\bibfnamefont{E.~D.~L.} \bibnamefont{Rienks}},
  \bibinfo{author}{\bibfnamefont{M.~F.} \bibnamefont{Jensen}},
  \bibnamefont{and} \bibinfo{author}{\bibfnamefont{P.}~\bibnamefont{Hofmann}},
  \bibinfo{journal}{Phys. Rev. Lett.} \textbf{\bibinfo{volume}{104}},
  \bibinfo{pages}{256803} (\bibinfo{year}{2010}).

\bibitem[{\citenamefont{Chiang}(2000)}]{Chiang:2000}
\bibinfo{author}{\bibfnamefont{T.~C.} \bibnamefont{Chiang}},
  \bibinfo{journal}{Surface Science Reports} \textbf{\bibinfo{volume}{39}},
  \bibinfo{pages}{181} (\bibinfo{year}{2000}), ISSN \bibinfo{issn}{0167-5729}.

\bibitem[{\citenamefont{Milun et~al.}(2002)\citenamefont{Milun, Pervan, and
  Woodruff}}]{Milun:2002}
\bibinfo{author}{\bibfnamefont{M.}~\bibnamefont{Milun}},
  \bibinfo{author}{\bibfnamefont{P.}~\bibnamefont{Pervan}}, \bibnamefont{and}
  \bibinfo{author}{\bibfnamefont{D.~P.} \bibnamefont{Woodruff}},
  \bibinfo{journal}{Reports on Progress in Physics}
  \textbf{\bibinfo{volume}{65}}, \bibinfo{pages}{99} (\bibinfo{year}{2002}).

\bibitem[{\citenamefont{Bastard}(1988)}]{Bastard:1988}
\bibinfo{author}{\bibfnamefont{G.}~\bibnamefont{Bastard}},
  \emph{\bibinfo{title}{Wave Mechanics Applied to Semiconductor
  Heterostructures}} (\bibinfo{publisher}{Wiley, New York},
  \bibinfo{year}{1988}).

\bibitem[{\citenamefont{Louie et~al.}(1980)\citenamefont{Louie, Thiry,
  Pinchaux, Petroff, Chanderis, and Lecante}}]{Louie:1980}
\bibinfo{author}{\bibfnamefont{S.~G.} \bibnamefont{Louie}},
  \bibinfo{author}{\bibfnamefont{P.}~\bibnamefont{Thiry}},
  \bibinfo{author}{\bibfnamefont{R.}~\bibnamefont{Pinchaux}},
  \bibinfo{author}{\bibfnamefont{Y.}~\bibnamefont{Petroff}},
  \bibinfo{author}{\bibfnamefont{D.}~\bibnamefont{Chanderis}},
  \bibnamefont{and} \bibinfo{author}{\bibfnamefont{J.}~\bibnamefont{Lecante}},
  \bibinfo{journal}{Physical Review Letters} \textbf{\bibinfo{volume}{44}},
  \bibinfo{pages}{549} (\bibinfo{year}{1980}).

\bibitem[{\citenamefont{Hofmann et~al.}(2002)\citenamefont{Hofmann,
  S{\o}ndergaard, Agergaard, Hoffmann, Gayone, Zampieri, Lizzit, and
  Baraldi}}]{Hofmann:2002}
\bibinfo{author}{\bibfnamefont{P.}~\bibnamefont{Hofmann}},
  \bibinfo{author}{\bibfnamefont{C.}~\bibnamefont{S{\o}ndergaard}},
  \bibinfo{author}{\bibfnamefont{S.}~\bibnamefont{Agergaard}},
  \bibinfo{author}{\bibfnamefont{S.~V.} \bibnamefont{Hoffmann}},
  \bibinfo{author}{\bibfnamefont{J.~E.} \bibnamefont{Gayone}},
  \bibinfo{author}{\bibfnamefont{G.}~\bibnamefont{Zampieri}},
  \bibinfo{author}{\bibfnamefont{S.}~\bibnamefont{Lizzit}}, \bibnamefont{and}
  \bibinfo{author}{\bibfnamefont{A.}~\bibnamefont{Baraldi}},
  \bibinfo{journal}{Physical Review B} \textbf{\bibinfo{volume}{66}},
  \bibinfo{pages}{245422} (\bibinfo{year}{2002}).

\bibitem[{\citenamefont{Navr\'atil et~al.}(2004)\citenamefont{Navr\'atil,
  Hor\'ak, Plech\'acek, Kamba, Lost'\'ak, Dyck, Chen, and
  Uher}}]{Navratil:2004}
\bibinfo{author}{\bibfnamefont{J.}~\bibnamefont{Navr\'atil}},
  \bibinfo{author}{\bibfnamefont{J.}~\bibnamefont{Hor\'ak}},
  \bibinfo{author}{\bibfnamefont{T.}~\bibnamefont{Plech\'acek}},
  \bibinfo{author}{\bibfnamefont{S.}~\bibnamefont{Kamba}},
  \bibinfo{author}{\bibfnamefont{P.}~\bibnamefont{Lost'\'ak}},
  \bibinfo{author}{\bibfnamefont{J.}~\bibnamefont{Dyck}},
  \bibinfo{author}{\bibfnamefont{W.}~\bibnamefont{Chen}}, \bibnamefont{and}
  \bibinfo{author}{\bibfnamefont{C.}~\bibnamefont{Uher}},
  \bibinfo{journal}{Journal of Solid State Chemistry}
  \textbf{\bibinfo{volume}{177}}, \bibinfo{pages}{1704} (\bibinfo{year}{2004}),
  ISSN \bibinfo{issn}{0022-4596}.

\bibitem[{\citenamefont{Hoffmann et~al.}(2004)\citenamefont{Hoffmann,
  S{\o}ndergaard, Schultz, Li, and Hofmann}}]{Hoffmann:2004}
\bibinfo{author}{\bibfnamefont{S.~V.} \bibnamefont{Hoffmann}},
  \bibinfo{author}{\bibfnamefont{C.}~\bibnamefont{S{\o}ndergaard}},
  \bibinfo{author}{\bibfnamefont{C.}~\bibnamefont{Schultz}},
  \bibinfo{author}{\bibfnamefont{Z.}~\bibnamefont{Li}}, \bibnamefont{and}
  \bibinfo{author}{\bibfnamefont{P.}~\bibnamefont{Hofmann}},
  \bibinfo{journal}{Nuclear Instruments and Methods in Physics Research, A}
  \textbf{\bibinfo{volume}{523}}, \bibinfo{pages}{441} (\bibinfo{year}{2004}).

\bibitem[{\citenamefont{Himpsel et~al.}(1983)\citenamefont{Himpsel, Hollinger,
  and Pollak}}]{Himpsel:1983b}
\bibinfo{author}{\bibfnamefont{F.~J.} \bibnamefont{Himpsel}},
  \bibinfo{author}{\bibfnamefont{G.}~\bibnamefont{Hollinger}},
  \bibnamefont{and} \bibinfo{author}{\bibfnamefont{R.~A.}
  \bibnamefont{Pollak}}, \bibinfo{journal}{Physical Review B}
  \textbf{\bibinfo{volume}{28}}, \bibinfo{pages}{7014} (\bibinfo{year}{1983}).

\bibitem[{\citenamefont{Kuroda et~al.}(2010)\citenamefont{Kuroda, Arita,
  Miyamoto, Ye, Jiang, Kimura, Krasovskii, Chulkov, Iwasawa, Okuda
  et~al.}}]{Kuroda:2010}
\bibinfo{author}{\bibfnamefont{K.}~\bibnamefont{Kuroda}},
  \bibinfo{author}{\bibfnamefont{M.}~\bibnamefont{Arita}},
  \bibinfo{author}{\bibfnamefont{K.}~\bibnamefont{Miyamoto}},
  \bibinfo{author}{\bibfnamefont{M.}~\bibnamefont{Ye}},
  \bibinfo{author}{\bibfnamefont{J.}~\bibnamefont{Jiang}},
  \bibinfo{author}{\bibfnamefont{A.}~\bibnamefont{Kimura}},
  \bibinfo{author}{\bibfnamefont{E.~E.} \bibnamefont{Krasovskii}},
  \bibinfo{author}{\bibfnamefont{E.~V.} \bibnamefont{Chulkov}},
  \bibinfo{author}{\bibfnamefont{H.}~\bibnamefont{Iwasawa}},
  \bibinfo{author}{\bibfnamefont{T.}~\bibnamefont{Okuda}},
  \bibnamefont{et~al.}, \bibinfo{journal}{Phys. Rev. Lett.}
  \textbf{\bibinfo{volume}{105}}, \bibinfo{pages}{076802}
  (\bibinfo{year}{2010}).

\bibitem[{\citenamefont{Fu}(2009)}]{Fu:2009}
\bibinfo{author}{\bibfnamefont{L.}~\bibnamefont{Fu}}, \bibinfo{journal}{Phys.
  Rev. Lett.} \textbf{\bibinfo{volume}{103}}, \bibinfo{pages}{266801}
  (\bibinfo{year}{2009}).

\bibitem[{\citenamefont{Hor et~al.}(2009)\citenamefont{Hor, Richardella,
  Roushan, Xia, Checkelsky, Yazdani, Hasan, Ong, and Cava}}]{Hor:2009}
\bibinfo{author}{\bibfnamefont{Y.~S.} \bibnamefont{Hor}},
  \bibinfo{author}{\bibfnamefont{A.}~\bibnamefont{Richardella}},
  \bibinfo{author}{\bibfnamefont{P.}~\bibnamefont{Roushan}},
  \bibinfo{author}{\bibfnamefont{Y.}~\bibnamefont{Xia}},
  \bibinfo{author}{\bibfnamefont{J.~G.} \bibnamefont{Checkelsky}},
  \bibinfo{author}{\bibfnamefont{A.}~\bibnamefont{Yazdani}},
  \bibinfo{author}{\bibfnamefont{M.~Z.} \bibnamefont{Hasan}},
  \bibinfo{author}{\bibfnamefont{N.~P.} \bibnamefont{Ong}}, \bibnamefont{and}
  \bibinfo{author}{\bibfnamefont{R.~J.} \bibnamefont{Cava}},
  \bibinfo{journal}{Phys. Rev. B} \textbf{\bibinfo{volume}{79}},
  \bibinfo{pages}{195208} (\bibinfo{year}{2009}).

\bibitem[{\citenamefont{Wyckoff}(1963)}]{Wyckoff:1963}
\bibinfo{author}{\bibfnamefont{R.~W.~G.} \bibnamefont{Wyckoff}},
  \emph{\bibinfo{title}{Crystal structures}}
  (\bibinfo{publisher}{Interscience}, \bibinfo{year}{1963}).

\bibitem[{\citenamefont{Hatch et~al.}(2011)\citenamefont{Hatch, Bianchi, Guan,
  Bao, Mi, Iversen, Nilsson, Hornek\ae{}r, and Hofmann}}]{Hatch:2011}
\bibinfo{author}{\bibfnamefont{R.~C.} \bibnamefont{Hatch}},
  \bibinfo{author}{\bibfnamefont{M.}~\bibnamefont{Bianchi}},
  \bibinfo{author}{\bibfnamefont{D.}~\bibnamefont{Guan}},
  \bibinfo{author}{\bibfnamefont{S.}~\bibnamefont{Bao}},
  \bibinfo{author}{\bibfnamefont{J.}~\bibnamefont{Mi}},
  \bibinfo{author}{\bibfnamefont{B.~B.} \bibnamefont{Iversen}},
  \bibinfo{author}{\bibfnamefont{L.}~\bibnamefont{Nilsson}},
  \bibinfo{author}{\bibfnamefont{L.}~\bibnamefont{Hornek\ae{}r}},
  \bibnamefont{and} \bibinfo{author}{\bibfnamefont{P.}~\bibnamefont{Hofmann}},
  \bibinfo{journal}{Phys. Rev. B} \textbf{\bibinfo{volume}{83}},
  \bibinfo{pages}{241303} (\bibinfo{year}{2011}).

\bibitem[{\citenamefont{Noguchi et~al.}(1991)\citenamefont{Noguchi, Hirakawa,
  and Ikoma}}]{Noguchi:1991}
\bibinfo{author}{\bibfnamefont{M.}~\bibnamefont{Noguchi}},
  \bibinfo{author}{\bibfnamefont{K.}~\bibnamefont{Hirakawa}}, \bibnamefont{and}
  \bibinfo{author}{\bibfnamefont{T.}~\bibnamefont{Ikoma}},
  \bibinfo{journal}{Physical Review Letters} \textbf{\bibinfo{volume}{66}},
  \bibinfo{pages}{2243} (\bibinfo{year}{1991}), \bibinfo{note}{cited By (since
  1996) 103},
  \urlprefix\url{http://www.scopus.com/inward/record.url?eid=2-s2.0-4243458390&partnerID=40&md5=6c911a3f56a78e62d6fc183812f02afa}.

\bibitem[{\citenamefont{Watanabe and Maeda}(1997)}]{Watanabe:1997}
\bibinfo{author}{\bibfnamefont{Y.}~\bibnamefont{Watanabe}} \bibnamefont{and}
  \bibinfo{author}{\bibfnamefont{F.}~\bibnamefont{Maeda}},
  \bibinfo{journal}{Applied Surface Science}
  \textbf{\bibinfo{volume}{117--118}}, \bibinfo{pages}{735 }
  (\bibinfo{year}{1997}), ISSN \bibinfo{issn}{0169-4332},
  \urlprefix\url{http://www.sciencedirect.com/science/article/pii/S0169433297801742}.

\bibitem[{\citenamefont{Deng et~al.}(2000)\citenamefont{Deng, Kwok, Lau, and
  Cao}}]{Deng:2000}
\bibinfo{author}{\bibfnamefont{Z.~W.} \bibnamefont{Deng}},
  \bibinfo{author}{\bibfnamefont{R.~W.~M.} \bibnamefont{Kwok}},
  \bibinfo{author}{\bibfnamefont{W.~M.} \bibnamefont{Lau}}, \bibnamefont{and}
  \bibinfo{author}{\bibfnamefont{L.~L.} \bibnamefont{Cao}},
  \bibinfo{journal}{Applied Surface Science} \textbf{\bibinfo{volume}{158}},
  \bibinfo{pages}{58} (\bibinfo{year}{2000}),
  \urlprefix\url{http://www.sciencedirect.com/science/article/B6THY-40F1R9X-B/2/5ee8320a48acedfe964e4e6ed059853a}.

\bibitem[{\citenamefont{Lowe et~al.}(2002)\citenamefont{Lowe, Veal, McConville,
  Bell, Tsukamoto, and Koguchi}}]{Lowe:2002}
\bibinfo{author}{\bibfnamefont{M.}~\bibnamefont{Lowe}},
  \bibinfo{author}{\bibfnamefont{T.}~\bibnamefont{Veal}},
  \bibinfo{author}{\bibfnamefont{C.}~\bibnamefont{McConville}},
  \bibinfo{author}{\bibfnamefont{G.}~\bibnamefont{Bell}},
  \bibinfo{author}{\bibfnamefont{S.}~\bibnamefont{Tsukamoto}},
  \bibnamefont{and} \bibinfo{author}{\bibfnamefont{N.}~\bibnamefont{Koguchi}},
  \bibinfo{journal}{Journal of Crystal Growth}
  \textbf{\bibinfo{volume}{237--239, Part 1}}, \bibinfo{pages}{196 }
  (\bibinfo{year}{2002}), ISSN \bibinfo{issn}{0022-0248},
  \bibinfo{note}{<ce:title>The thirteenth international conference on Crystal
  Growth in conj unction with the eleventh international conference on Vapor
  Growth and Epitaxy</ce:title>},
  \urlprefix\url{http://www.sciencedirect.com/science/article/pii/S0022024801018991}.

\bibitem[{\citenamefont{Chen et~al.}(2012)\citenamefont{Chen, He, Weng, Zhang,
  Zhao, Liu, Jia, Mou, Liu, He et~al.}}]{Chen:2012}
\bibinfo{author}{\bibfnamefont{C.}~\bibnamefont{Chen}},
  \bibinfo{author}{\bibfnamefont{S.}~\bibnamefont{He}},
  \bibinfo{author}{\bibfnamefont{H.}~\bibnamefont{Weng}},
  \bibinfo{author}{\bibfnamefont{W.}~\bibnamefont{Zhang}},
  \bibinfo{author}{\bibfnamefont{L.}~\bibnamefont{Zhao}},
  \bibinfo{author}{\bibfnamefont{H.}~\bibnamefont{Liu}},
  \bibinfo{author}{\bibfnamefont{X.}~\bibnamefont{Jia}},
  \bibinfo{author}{\bibfnamefont{D.}~\bibnamefont{Mou}},
  \bibinfo{author}{\bibfnamefont{S.}~\bibnamefont{Liu}},
  \bibinfo{author}{\bibfnamefont{J.}~\bibnamefont{He}}, \bibnamefont{et~al.},
  \bibinfo{journal}{Proceedings of the National Academy of Sciences}
  (\bibinfo{year}{2012}),
  \urlprefix\url{http://www.pnas.org/content/early/2012/02/17/1115555109.abstract}.

\bibitem[{\citenamefont{Bihlmayer et~al.}(2007)\citenamefont{Bihlmayer, Blugel,
  and Chulkov}}]{Bihlmayer:2007}
\bibinfo{author}{\bibfnamefont{G.}~\bibnamefont{Bihlmayer}},
  \bibinfo{author}{\bibfnamefont{S.}~\bibnamefont{Blugel}}, \bibnamefont{and}
  \bibinfo{author}{\bibfnamefont{E.~V.} \bibnamefont{Chulkov}},
  \bibinfo{journal}{Physical Review B} \textbf{\bibinfo{volume}{75}},
  \bibinfo{eid}{195414} (pages~\bibinfo{numpages}{6}) (\bibinfo{year}{2007}),
  \urlprefix\url{http://link.aps.org/abstract/PRB/v75/e195414}.

\bibitem[{\citenamefont{Bentmann et~al.}(2011)\citenamefont{Bentmann, Kuzumaki,
  Bihlmayer, Bl\"ugel, Chulkov, Reinert, and Sakamoto}}]{Bentmann:2011}
\bibinfo{author}{\bibfnamefont{H.}~\bibnamefont{Bentmann}},
  \bibinfo{author}{\bibfnamefont{T.}~\bibnamefont{Kuzumaki}},
  \bibinfo{author}{\bibfnamefont{G.}~\bibnamefont{Bihlmayer}},
  \bibinfo{author}{\bibfnamefont{S.}~\bibnamefont{Bl\"ugel}},
  \bibinfo{author}{\bibfnamefont{E.~V.} \bibnamefont{Chulkov}},
  \bibinfo{author}{\bibfnamefont{F.}~\bibnamefont{Reinert}}, \bibnamefont{and}
  \bibinfo{author}{\bibfnamefont{K.}~\bibnamefont{Sakamoto}},
  \bibinfo{journal}{Phys. Rev. B} \textbf{\bibinfo{volume}{84}},
  \bibinfo{pages}{115426} (\bibinfo{year}{2011}),
  \urlprefix\url{http://link.aps.org/doi/10.1103/PhysRevB.84.115426}.

\bibitem[{\citenamefont{Koleini et~al.}(2011)\citenamefont{Koleini, Frauenheim,
  and Yan}}]{Koleini:ARXIV}
\bibinfo{author}{\bibfnamefont{M.}~\bibnamefont{Koleini}},
  \bibinfo{author}{\bibfnamefont{T.}~\bibnamefont{Frauenheim}},
  \bibnamefont{and} \bibinfo{author}{\bibfnamefont{B.}~\bibnamefont{Yan}},
  \bibinfo{journal}{ArXiv e-prints} \textbf{\bibinfo{volume}{1109.4000}}
  (\bibinfo{year}{2011}), \eprint{1109.4000}.

\bibitem[{\citenamefont{Zhang et~al.}(2010)\citenamefont{Zhang, Yu, Zhang, Dai,
  and Fang}}]{Zhang:2010}
\bibinfo{author}{\bibfnamefont{W.}~\bibnamefont{Zhang}},
  \bibinfo{author}{\bibfnamefont{R.}~\bibnamefont{Yu}},
  \bibinfo{author}{\bibfnamefont{H.-J.} \bibnamefont{Zhang}},
  \bibinfo{author}{\bibfnamefont{X.}~\bibnamefont{Dai}}, \bibnamefont{and}
  \bibinfo{author}{\bibfnamefont{Z.}~\bibnamefont{Fang}}, \bibinfo{journal}{New
  Journal of Physics} \textbf{\bibinfo{volume}{12}}, \bibinfo{pages}{065013}
  (\bibinfo{year}{2010}),
  \urlprefix\url{http://stacks.iop.org/1367-2630/12/i=6/a=065013}.

\bibitem[{\citenamefont{Paraskevopoulos
  et~al.}(1988)\citenamefont{Paraskevopoulos, Hatzikraniotis, Chrisafis,
  Zamani, Stoemenos, Economou, Alexiadis, and
  Balkanski}}]{Paraskevopoulos:1988}
\bibinfo{author}{\bibfnamefont{K.}~\bibnamefont{Paraskevopoulos}},
  \bibinfo{author}{\bibfnamefont{E.}~\bibnamefont{Hatzikraniotis}},
  \bibinfo{author}{\bibfnamefont{K.}~\bibnamefont{Chrisafis}},
  \bibinfo{author}{\bibfnamefont{M.}~\bibnamefont{Zamani}},
  \bibinfo{author}{\bibfnamefont{J.}~\bibnamefont{Stoemenos}},
  \bibinfo{author}{\bibfnamefont{N.~A.} \bibnamefont{Economou}},
  \bibinfo{author}{\bibfnamefont{K.}~\bibnamefont{Alexiadis}},
  \bibnamefont{and}
  \bibinfo{author}{\bibfnamefont{M.}~\bibnamefont{Balkanski}},
  \bibinfo{journal}{Materials Science and Engineering B}
  \textbf{\bibinfo{volume}{1}}, \bibinfo{pages}{147} (\bibinfo{year}{1988}),
  \urlprefix\url{http://www.scopus.com/inward/record.url?eid=2-s2.0-0024106946&partnerID=40&md5=bf76ba4bc71f08470cfe7ec4ce671d5c}.

\bibitem[{\citenamefont{Hor et~al.}(2010)\citenamefont{Hor, Williams,
  Checkelsky, Roushan, Seo, Xu, Zandbergen, Yazdani, Ong, and
  Cava}}]{Hor:2010b}
\bibinfo{author}{\bibfnamefont{Y.~S.} \bibnamefont{Hor}},
  \bibinfo{author}{\bibfnamefont{A.~J.} \bibnamefont{Williams}},
  \bibinfo{author}{\bibfnamefont{J.~G.} \bibnamefont{Checkelsky}},
  \bibinfo{author}{\bibfnamefont{P.}~\bibnamefont{Roushan}},
  \bibinfo{author}{\bibfnamefont{J.}~\bibnamefont{Seo}},
  \bibinfo{author}{\bibfnamefont{Q.}~\bibnamefont{Xu}},
  \bibinfo{author}{\bibfnamefont{H.~W.} \bibnamefont{Zandbergen}},
  \bibinfo{author}{\bibfnamefont{A.}~\bibnamefont{Yazdani}},
  \bibinfo{author}{\bibfnamefont{N.~P.} \bibnamefont{Ong}}, \bibnamefont{and}
  \bibinfo{author}{\bibfnamefont{R.~J.} \bibnamefont{Cava}},
  \bibinfo{journal}{Phys. Rev. Lett.} \textbf{\bibinfo{volume}{104}},
  \bibinfo{pages}{057001} (\bibinfo{year}{2010}).

\bibitem[{\citenamefont{Yazyev et~al.}(2012)\citenamefont{Yazyev, Kioupakis,
  Moore, and Louie}}]{Yazyev:2012}
\bibinfo{author}{\bibfnamefont{O.~V.} \bibnamefont{Yazyev}},
  \bibinfo{author}{\bibfnamefont{E.}~\bibnamefont{Kioupakis}},
  \bibinfo{author}{\bibfnamefont{J.~E.} \bibnamefont{Moore}}, \bibnamefont{and}
  \bibinfo{author}{\bibfnamefont{S.~G.} \bibnamefont{Louie}},
  \bibinfo{journal}{Phys. Rev. B} \textbf{\bibinfo{volume}{85}},
  \bibinfo{pages}{161101} (\bibinfo{year}{2012}),
  \urlprefix\url{http://link.aps.org/doi/10.1103/PhysRevB.85.161101}.

\end{thebibliography}
\end{document}